\documentclass[a4paper]{article}
\pdfoutput=1
\usepackage{jheppub}
\usepackage{booktabs}
\usepackage{multirow}
\usepackage{graphicx}
\usepackage{epsfig}
\usepackage{lscape}
\usepackage{rotating}
\usepackage{hyperref}
\usepackage{amsmath}
\usepackage{lineno}
\usepackage{hyperref}
\usepackage{graphicx}
\usepackage{hyperref}
\usepackage{natbib}
\usepackage{subcaption}
\usepackage{epstopdf}
\usepackage{url}
\usepackage[utf8]{inputenc}
\usepackage{amsmath,amssymb,verbatim,mathrsfs}
\usepackage{textcomp}

\graphicspath{{Figs/}}

\title{Vector-like quarks coupling discrimination at the LHC and future hadron colliders}

\author[1]{D. Barducci}
\author[2,3]{and L. Panizzi}

\emailAdd{daniele.barducci@sissa.it}
\emailAdd{panizzi@ge.infn.it}

\affiliation[1]{SISSA and INFN, Sezione di Trieste, via Bonomea 265, 34136 Trieste, Italy}
\affiliation[2]{Dipartimento di Fisica, Universit\`a di Genova and INFN, Sezione di Genova,via Dodecaneso 33, 16146 Genova, Italy}
\affiliation[3]{School of Physics and Astronomy, University of Southampton, Highfield, Southampton SO17 1BJ, UK}

\begin{document}

\abstract{The existence of new coloured states with spin one-half, {\it i.e.} extra-quarks, is a striking prediction of various classes of new physics models. Should one of these states be discovered during the 13 TeV runs of the LHC or at future high energy hadron colliders, understanding its properties will be crucial in order to shed light on the underlying model structure.
Depending on the extra-quarks quantum number under $SU(2)_{L}$, their coupling to Standard Model quarks and bosons have either a dominant left- or right-handed chiral component.
By exploiting the polarisation properties of the top quarks arising from the decay of pair-produced extra quarks, we show how it is possible to discriminate among the two hypothesis in the whole discovery range currently accessible at the LHC, thus effectively narrowing down the possible interpretations of a discovered state in terms of new physics scenarios. Moreover, we estimate the discovery and discrimination power of future prototype hadron colliders with centre of mass energies of 33 and 100 TeV.
}

\begin{flushright}
\hspace{3cm} 
SISSA-47-2017-FISI
\end{flushright}

\maketitle

\section{Introduction}

Multiple evidences of the existence of New Physics (NP) beyond the Standard Model (BSM) -- such as the observation of neutrino oscillations, the matter-antimatter asymmetry present in the Universe and Dark Matter phenomena --
together with the search for a mechanism stabilising the electroweak (EW) scale to its measured value, 
 have been  the driving force of an extraordinary effort by the theoretical community in the development of new theoretical ideas which extend the SM and provide an explanation to such observations. Such models usually predict, in turn, the existence of new states, whose masses can generally be within the reach of current or future colliders. 
The experimental collaborations at the Large Hadron Collider (LHC) are currently pursuing a large number of searches, aimed at discovering signals generated by such new states in a wide range of signatures. However, the chase for NP has been so far surprisingly unsuccessful. Therefore, bounds on the masses of new particles are pushed to higher and higher values as the collected luminosity increases and more data are collected. Despite the current lack of hints about where NP could (and hopefully will) be found, it is crucial to be prepared in case a discovery is made: if a new state is observed at the LHC, the determination of its properties (such as mass, width, spin and couplings) would be extremely important for embedding such state in theoretically motivated BSM scenarios and possibly rule out whole classes of models.

A certain number of hypothetical new particles have been searched with particular interest, due to their large discovery potential; vector-like quarks (VLQs) are among those states and they are the subject of our analysis.
VLQs are heavy quarks whose left- and right-handed chiral components transform under the same representation of the SM gauge group. They appear in a variety of BSM scenarios formulated to seek for an answer to the problem of the naturalness of the EW scale. 
These states are for example naturally present in Composite Higgs models (CHMs)~\cite{Kaplan:1983fs} where the top quark Yukawa coupling is generated via the partial compositness mechanism~\cite{Kaplan:1991dc} and are for this reason also know as top partners. Due to the crucial role that they play in controlling the level of fine tuning of CHMs, they have been object of intensive experimental searches at the LHC. 
VLQs also appear in other BSM scenarios, as in models of extra dimensions as Kaluza-Klein excitation of the SM fermions~\cite{Antoniadis:1990ew,Csaki:2003sh,Cacciapaglia:2009pa}, models aiming at generating the fermions masses~\cite{Grossman:1999ra}, model with extended gauge symmetries like little Higgs models~\cite{ArkaniHamed:2002qx} and others~\cite{Abbas:2017hzw,Abbas:2017vle}.
In specific NP realisations, VLQs can have different charge assignments under the SM EW gauge group $SU(2)_L \times U(1)_Y$. Hence there exists the possibility of having multiple VLQs, also with exotic electric charges, present at a mass scale accessible at the LHC, therefore providing a rich phenomenology~\cite{AguilarSaavedra:2009es,DeSimone:2012fs,Aguilar-Saavedra:2013qpa,Barducci:2013zaa,Alok:2014yua,Barducci:2014ila,Cacciapaglia:2015ixa,Alok:2015iha}.

The minimal scenarios which present VLQs in addition to the SM particle content are those in which the VLQs interact with the SM quarks via Yukawa-type interactions. By classifying the VLQs representations through their $SU(2)_L$ quantum numbers, gauge invariant renormalizable operators can be written only for singlets, doublets and triplets representations~\cite{delAguila:2000rc,Okada:2012gy}. 
However, a remarkable property of VLQs in extensions of the SM where they are the only new states is that regardless of the number of VLQs and of their mixing patterns with the SM quarks and between themselves, the couplings of the VLQs with the SM bosons and quarks always have a dominant chiral component; this depends only on whether their weak isospin is integer or half-integer, the other component being suppressed by a factor proportional to $m_{q^{\rm SM}}/m_{VLQ}$, with $m_{q^{\rm SM}}$ the mass of the SM quark with which the VLQ mixes~\cite{delAguila:2000rc,Buchkremer:2013bha}. 
This property affects the polarisation of the gauge bosons and quarks arising from the VLQs decay. While, as we will show, the gauge bosons tend to have a dominant longitudinal polarisation, thus making them insensitive to the chiral structure of the interaction vertex, the polarisation of the final state quarks allows to extract useful information. 
In particular if the VLQs decay into a top quark, which decays before hadronisation takes place, its polarisation properties will affect the  kinematic distributions of the final decay products.
The modifications of the final state objects kinematics can slightly affect the reach of NP searches but, more importantly, in the fortunate event of a signal excess being observed, these differences can be used to probe the structure of the interactions between the VLQs and the SM sector~\footnote{See also~\cite{Godbole:2010kr,Belanger:2012tm} for related studies in the context of different NP models.}.

In this paper we will focus on the illustrative examples of VLQs with charges 2/3 and 5/3 interacting with the third generation of SM quarks, and in particular with the top quark, and we will analyse their pair production signatures to evaluate the possibility to discriminate between different assumptions about their couplings, and, as a consequence, about their representations under the SM $SU(2)_L$ gauge group. We will show how, in case a VLQ is discovered at the LHC, simple kinematics considerations can be used to determine the chiral structure of the VLQ coupling, thus effectively helping to narrow down the possible interpretation in terms of NP models. Furthermore we will estimate the reach for discovering (or excluding) VLQs and determining their properties for the next generation of prototype hadron colliders, with a centre of mass energy of 33 and 100 TeV.

The paper is organized as follows. In Sec.~\ref{sec:param} we parametrize the VLQs interaction in terms of a phenomenological lagrangian and discuss the polarisation properties of the VLQs decay products. 
In Sec.~\ref{sec:LHC} we will show the discrimination power of the LHC comparing it with its discovery reach, while in Sec.~\ref{sec:FCC} we will do the same for future high energy collider prototypes.
We then conclude in Sec.~\ref{sec:conc}.

\section{Parametrisation and polarisation properties}
\label{sec:param} 

In the minimal scenario where the VLQs interact with the SM sector only through Yukawa-type couplings,
gauge invariant renormalizable interactions can only be written for the restricted combinations of $SU(2)_L\times U(1)_Y$ quantum numbers shown in Tab.~\ref{tab:reppr}, where the $U(1)_{\rm em}$ charges of the $X,\;T,\;B$ and $Y$ VLQs read 5/3, 2/3, -1/3 and -4/3 respectively. After the Higgs gets a vacuum expectation value the VLQs mix with the SM quarks, with different rotations angles for their left- and right-handed chiral components. In particular one obtains the following relations between the mixing angles relating the interaction and mass eigenstates~\cite{Cacciapaglia:2010vn,Chen:2017hak}
\begin{equation}
\begin{split}
\frac{\tan{\theta^R}}{\tan{\theta^L}} = \frac{m_q^{SM}}{m_{VLQ}} \quad{\textrm{for}}\quad SU(2)_L =1,3 \qquad\qquad
\frac{\tan{\theta^L}}{\tan{\theta^R}} = \frac{m_q^{SM}}{m_{VLQ}} \quad{\textrm{for}}\quad SU(2)_L =2 \\
\end{split}
\end{equation}
where $m_q^{SM}$ refers to the SM quark which mixes with the considered VLQ.
Given the current bounds on VLQs masses which are of the order of 1 TeV (see {\emph{e.g.}}~\cite{Aaboud:2017qpr,CMS:2017wwc}), a hierarchy is obtained between the two mixing angles which, in the mass eigenbasis, reflects on a dominant chiral structure in the coupling between a VLQ, a SM quark and a SM boson. This property can be generalised to more complex extensions of the SM, which contain any number of VLQs which interact with the SM and between them through Yukawa-type couplings: the dominant coupling chirality of a VLQ is always related to its representation under $SU(2)_L$. If the VLQ belongs to a representation with integer weak isospin its couplings have a dominant left-handed chirality, while for half-integer weak isospins the couplings are dominantly right-handed~\cite{delAguila:2000rc,Buchkremer:2013bha}. 

\begin{table}[htbp]
\centering
\begin{tabular}{c|c|c|c}
$SU(2)_L$ 						& $U(1)_Y$ & $\psi$     & $L_y$  \\ 
\hline
\hline
\multirow{ 2}{*}{1} 				& 2/3		 & $T$  		  & $\bar q_L H^c t_R$  \\
										& -1/3 		 & $B$  		  & $\bar q_L H b_R$   \\ 
							\hline
\multirow{ 3}{*}{2}				& 7/6  		 & $(X,T)$    & $\bar \psi_L H t_R$  \\
										& 1/6 		 & $(T,B)$    & $\bar \psi_L H^c t_R$, $\bar \psi_L H b_R$  \\
										& -5/6 		 & $(B,Y)$    & $\bar \psi_L H^c b_R$  \\			
\hline
\multirow{ 2}{*}{3} 				&  2/3		 & $(X,T,B)$ & $\bar q_L\tau^a H^c \psi_R^a$\\		
										&  -1/3		 & $(T,B,Y)$ & $\bar q_L\tau^a H \psi_R^a$\\			
\end{tabular}
\caption{VLQs multiplet quantum numbers under $SU(2)_L$ and $U(1)_Y$ that allows to write gauge invariant renormalizable interactions with the SM sector through Yukawa-type terms. The $U(1)_{\rm em}$ quantum numbers of the $X,\;T,\;B$ and $Y$ VLQs are 5/3, 2/3, -1/3 and -4/3 respectively.}
\label{tab:reppr}
\end{table}

It is thus interesting to ask the following question: if a signal originated by a VLQ decaying into SM states is observed at the LHC, is it possible to understand the chiral structure of the coupling responsible for its decay and thus to restrict the possible $SU(2)_L$ representations to which it may belong? In the following we will restrict our study to VLQs decaying to SM top quarks. With this assumption, the problem can be addressed in two ways. Firstly, by studying the polarisation properties of the gauge boson arising from the VLQ decay, in analogy to what is done for the $t\to W^+ b$ process in the SM, and secondly by analysing the polarisation properties of the top quark. 

\subsection{Gauge boson polarisation}

\begin{figure}[!htbp]
\centering
\includegraphics[width=0.48\textwidth]{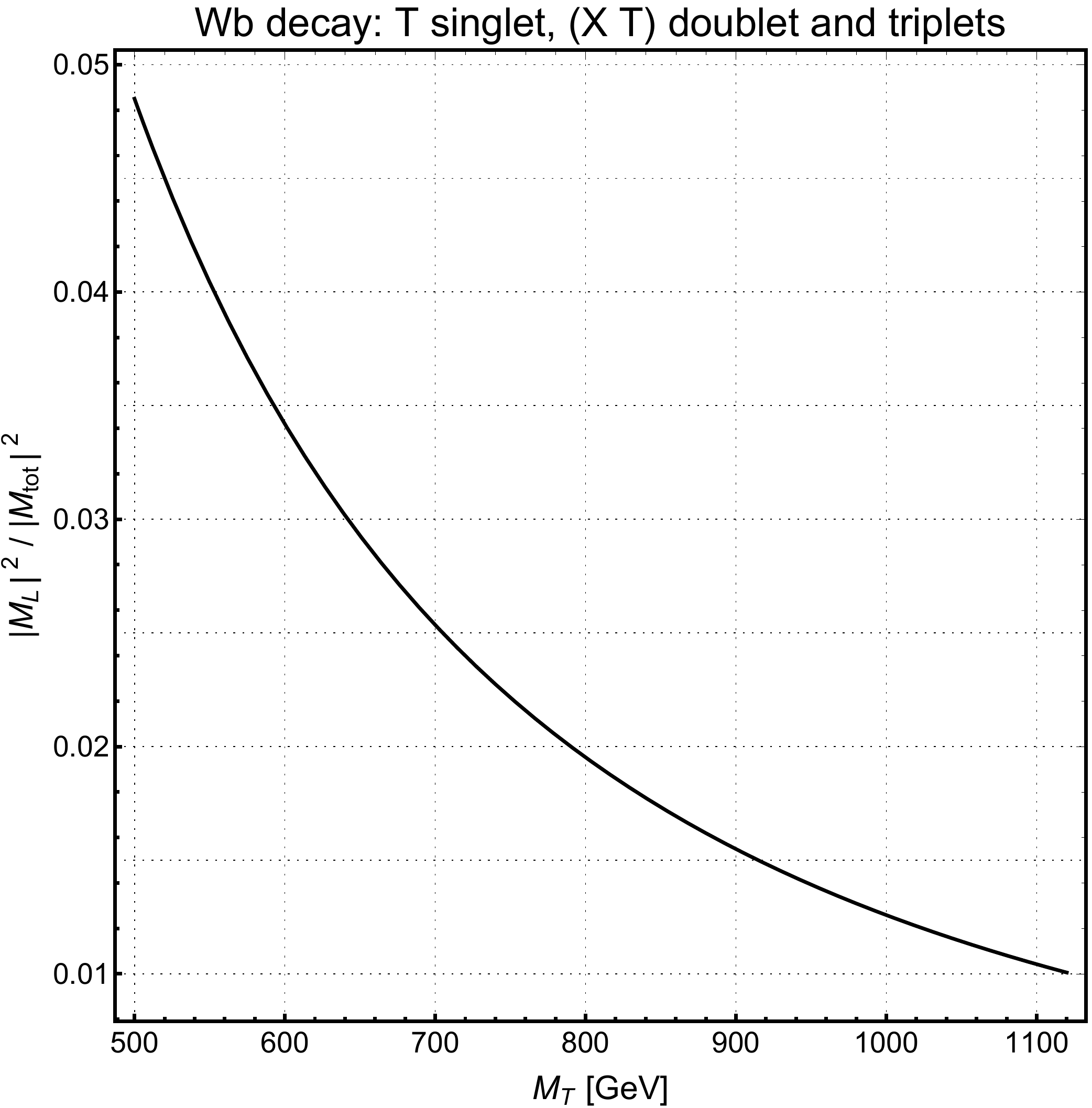}
\includegraphics[width=0.48\textwidth]{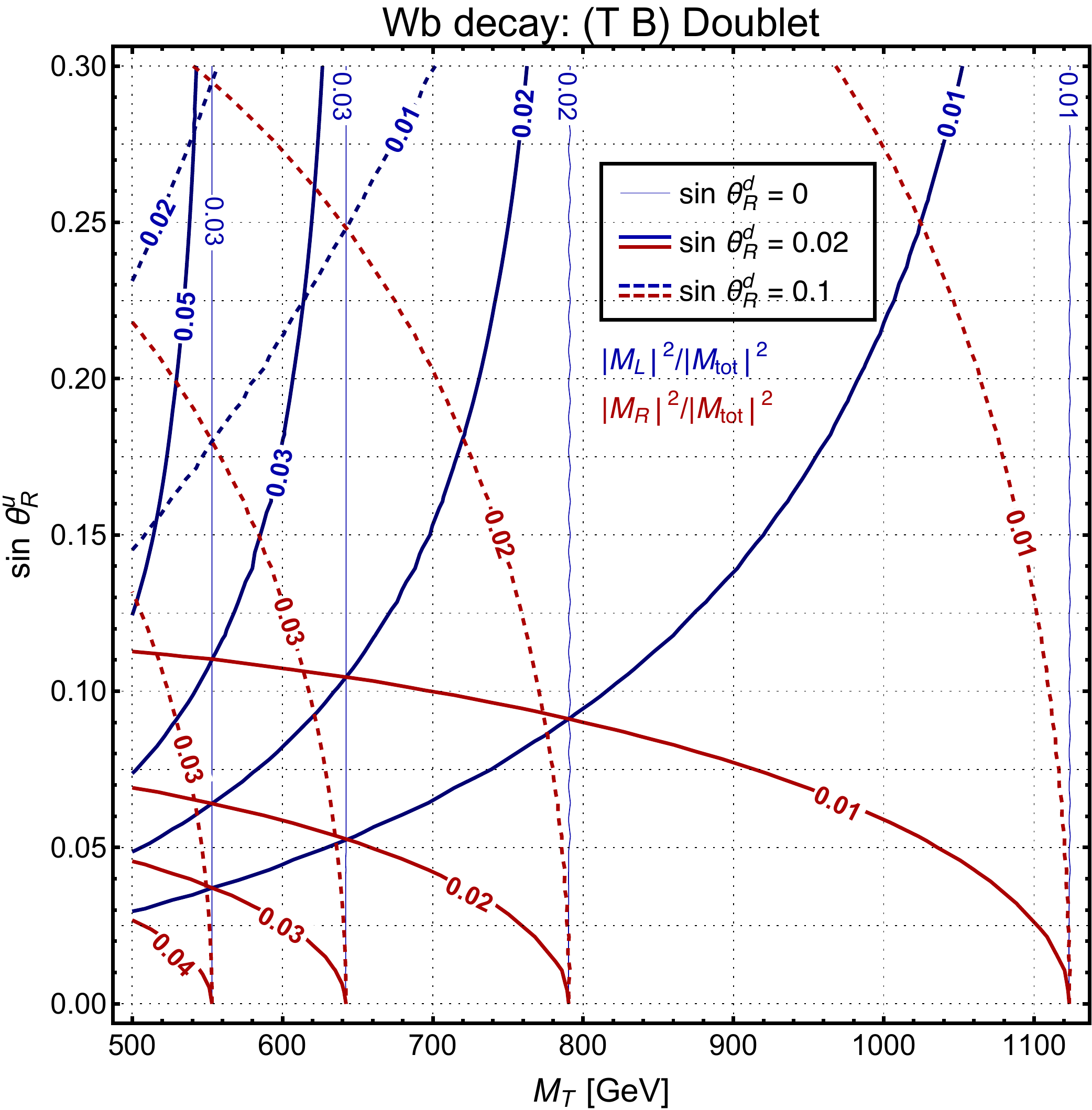}
\caption{Fraction of left-handed and right-handed polarised $W$ boson arising from the charged current decay of a $T$ VLQ. In the left panel the singlet, $(X~T)$ doublet and triplet representations are shown as function of $m_T$, while in the right panel the $(T~B)$ doublet representation is shown in the $m_T-\sin\theta^u_R$ plane for different values of $\sin\theta^d_R$.}
\label{fig:wpol}
\end{figure}

For the SM $t \to W^+ b$ process the polarisation fractions of the $W$ boson arising from the top quark decay can be extracted from a measurement of the angular distribution of the decay products of the top quark given by~\cite{Schroder:2014gca}
\begin{equation}
\frac{1}{\sigma} \frac{d\sigma}{d \cos \theta^*}=\frac{3}{4}(1-\cos^2\theta^*)F_0+\frac{3}{8}(1-\cos\theta^*)^2F_L+\frac{3}{8}(1+\cos\theta^*)^2F_R
\label{eq:dsigma}
\end{equation}
with $\theta^*$ the angle between the $W$ boson momentum in the top quark rest frame and the momentum of the down-type decay fermion in the rest frame of the $W$ boson. 
The SM expectation for the longitudinal, left and right-handed polarisation fractions together with the values measured by the CMS collaboration through pair production~\cite{Khachatryan:2016fky} and single production~\cite{Khachatryan:2014vma} processes are reported in Tab.~\ref{tab:WpolarisationSMtop}. We note in particular that through top quark pair production events, the longitudinal and left-handed polarisation fractions are determined with $\sim 5\%$ uncertainties which are already systematics-dominated.
\begin{table}
\centering
\small
\begin{tabular}{c|cc|cc}
\toprule
Polarisation & \multicolumn{2}{c|}{Pair production} & \multicolumn{2}{c}{Single production} \\
             & observed & SM expectation & observed & SM expectation \\
\midrule
$F_0$ & $0.681\pm0.012\pm0.023$ & $0.687\pm0.005$ & $0.720\pm0.039\pm0.037$ & $0.687\pm0.005$ \\
$F_L$ & $0.323\pm0.008\pm0.014$ & $0.311\pm0.005$ & $0.298\pm0.028\pm0.032$ & $0.311\pm0.005$ \\
$F_R$ & $-0.004\pm0.005\pm0.014$ & $0.0017\pm0.0001$ & $-0.018\pm0.019\pm0.011$ & $0.0017\pm0.0001$ \\
\bottomrule
\end{tabular}
\caption{\label{tab:WpolarisationSMtop} $W$ polarisations fractions from top decay measured at CMS for processes of pair~\cite{Khachatryan:2016fky} and single top~\cite{Khachatryan:2014vma} production at the 8 TeV LHC, together with the SM expectation~\cite{Czarnecki:2010gb}. The quoted experimental uncertainties correspond to the statistical and systematic errors respectively.}
\end{table}
We now want to see what are the polarisation fractions for the SM gauge bosons arising from the decay of a VLQ. We consider as an example the case of the $T$ VLQ which can decay into a $Wb$, $Zt$ and $Ht$ final state. Obviously, no information on the structure of the VLQ coupling can be inferred from the Higgs boson in the $Ht$ final state. 
\begin{table}[!htbp]
\centering
\footnotesize
\hspace*{-2pt}\begin{tabular}{c|l}
\toprule
& \multicolumn{1}{c}{Squared amplitudes for gauge boson polarisations} \\
\midrule
\midrule
\multicolumn{2}{c}{$W b$ decay}\\
\midrule
\midrule
\multirow{3}{*}{\begin{tabular}{c} $T$ \\ $(X~T)$ \end{tabular}} & 
  $|M|^2_L = \frac{g^2}{2} \sin^2\theta^u_L (m_T^2-m_W^2)$ \\
& $|M|^2_R = 0$ \\
& $|M|^2_0 = \frac{g^2}{4} \frac{m_T^2}{m_W^2} \sin^2\theta^u_L \left(m_T^2-m_W^2 \right) $ \\
\midrule
\multirow{3}{*}{$(T~B)$} &
  $|M|^2_L = \frac{g^2}{2} \sin^2\theta^u_R \frac{m_t^2(m_t^2-m_W^2)}{\cos^2\theta^u_R m_T^2+ \sin^2\theta^u_R m_t^2}$\\
  & $|M|^2_R = \frac{g^2}{2} \cos^2\theta^u_R \sin^2\theta^d_R (m_T^2-m_W^2) $ \\
& $\begin{array}{ll} |M|^2_0 = & \frac{g^2}{4} \frac{m_T^2}{m_W^2} \frac{\cos^4\theta^u_R \sin^2\theta^d_R m_T^2 +\sin^2\theta^u_R m_t^2 (1+ \cos^2\theta^u_R \sin^2\theta^d_R)}{\cos^2\theta^u_R m_T^2 + \sin^2\theta^u_R m_t^2} (m_T^2-m_W^2) \end{array}$ \\
\midrule
\multirow{3}{*}{\begin{tabular}{c} $(X~T~B)$ \\ $(T~B~Y)$ \end{tabular}} &
  $|M|^2_L = \frac{g^2}{2} (\sin\theta^u_L\cos\theta^d_L-\sqrt{2}\cos\theta^u_L\sin\theta^d_L)^2 (m_T^2-m_W^2) $ \\
& $|M|^2_R = 0 $ \\
& $|M|^2_0 = \frac{g^2}{4}\frac{m_T^2}{m_W^2} (\sin\theta^u_L\cos\theta^d_L-\sqrt{2}\cos\theta^u_L\sin\theta^d_L)^2 (m_T^2-m_W^2) $ \\
\midrule
\midrule
\multicolumn{2}{c}{$Z t$ decay}\\
\midrule
\midrule
\multirow{3}{*}{\begin{tabular}{c} $T$ \\ $(X~T~B)$ \end{tabular}} & 
  $|M|^2_L = \frac{g^2}{4 c_w^2} (\cos\theta^u_L \sin\theta^u_L)^2  (m_T^2 - m_Z^2 + m_t^2) $\\
& $|M|^2_R = 0$ \\
& $|M|^2_0 = \frac{g^2}{8 c_w^2} \frac{m_T^2}{m_Z^2} (\cos\theta^u_L \sin\theta^u_L)^2 \left(m_T^2 - m_Z^2 + m_t^2 ( \frac{m_t^2-m_Z^2}{m_T^2} - 2 ) \right)$ \\
\midrule
\multirow{3}{*}{$(X~T)$} &
  $|M|^2_L = \frac{g^2}{4 cw^2} (\cos\theta^u_R \sin\theta^u_R)^2 \frac{4 m_t^2 m_T^2 (m_T^2 - m_Z^2 + m_t^2)}{(m_T^2 \cos^2\theta^u_R + m_t^2 \sin^2\theta^u_R)^2} $ \\
& $|M|^2_R = \frac{g^2}{4 c_w^2} (\cos\theta^u_R \sin\theta^u_R)^2 (m_T^2 - m_Z^2 + m_t^2) $ \\
& $|M|^2_0 = \frac{g^2}{8 c_w^2} \frac{m_T^2}{m_Z^2} (\cos\theta^u_R \sin\theta^u_R)^2 \left(m_T^2 - m_Z^2 + m_t^2 (\frac{m_t^2-m_Z^2}{m_T^2} - 2 )\right) \left( \frac{4 m_t^2 m_T^2}{(m_T^2 \cos^2\theta^u_R + m_t^2 \sin^2\theta^u_R)^2} + 1 \right)$ \\
\bottomrule
\end{tabular}
\caption{\label{tab:Vpolarisation} 
Contribution from different $W$ and $Z$ bosons polarisations to the squared amplitude for the processes $T\to Wb$ an $T\to Zt$ in the limit $m_b\to 0$.
For the $Z$ boson, the $(T~B)$ doublet expressions can be obtained from the $T$ and $(X~T~B)$ ones, exchanging the left and right components; analogously, the $(T~B~Y)$ triplet expressions can be obtained from the $(X~T)$ doublet with the same exchange.}
\end{table}
We thus focus onto the other two decay modes computing the polarisation fractions of the SM gauge boson arising from the decay of a VLQ. The contribution from the different polarisations of the $W$ and $Z$ bosons to the squared amplitude in the limit $m_b\to 0$ are reported in Tab.~\ref{tab:Vpolarisation} for all the representations in which the $T$ can be embedded.
For the $Wb$ decay, in the limit $m_b\to0$, the longitudinal component reads in all cases $|M|^2_0 = \frac{1}{2} \frac{m_T^2}{m_W^2} ( |M|^2_L + |M|^2_R )$. It is also interesting to notice that for the singlet, $(X~T)$ doublet and the triplets representations the right-handed polarisation gives no contribution, and therefore, the ratio between the left-handed and total contributions is independent of the mixing angle and scales as $m_T^2/2m_W^2$. For the SM-like doublet representation the picture is more complicate, due to the presence of the right-handed contribution and the role of the mixing angle in the down sector. We show in Fig.~\ref{fig:wpol} the left and right polarisation fractions for all the choices of $SU(2)_L$ quantum numbers of the  $T$ VLQ. For the doublet we fix $\sin\theta^d_R$ to representative values. 
\begin{figure}[!htbp]
\centering
\includegraphics[width=0.48\textwidth]{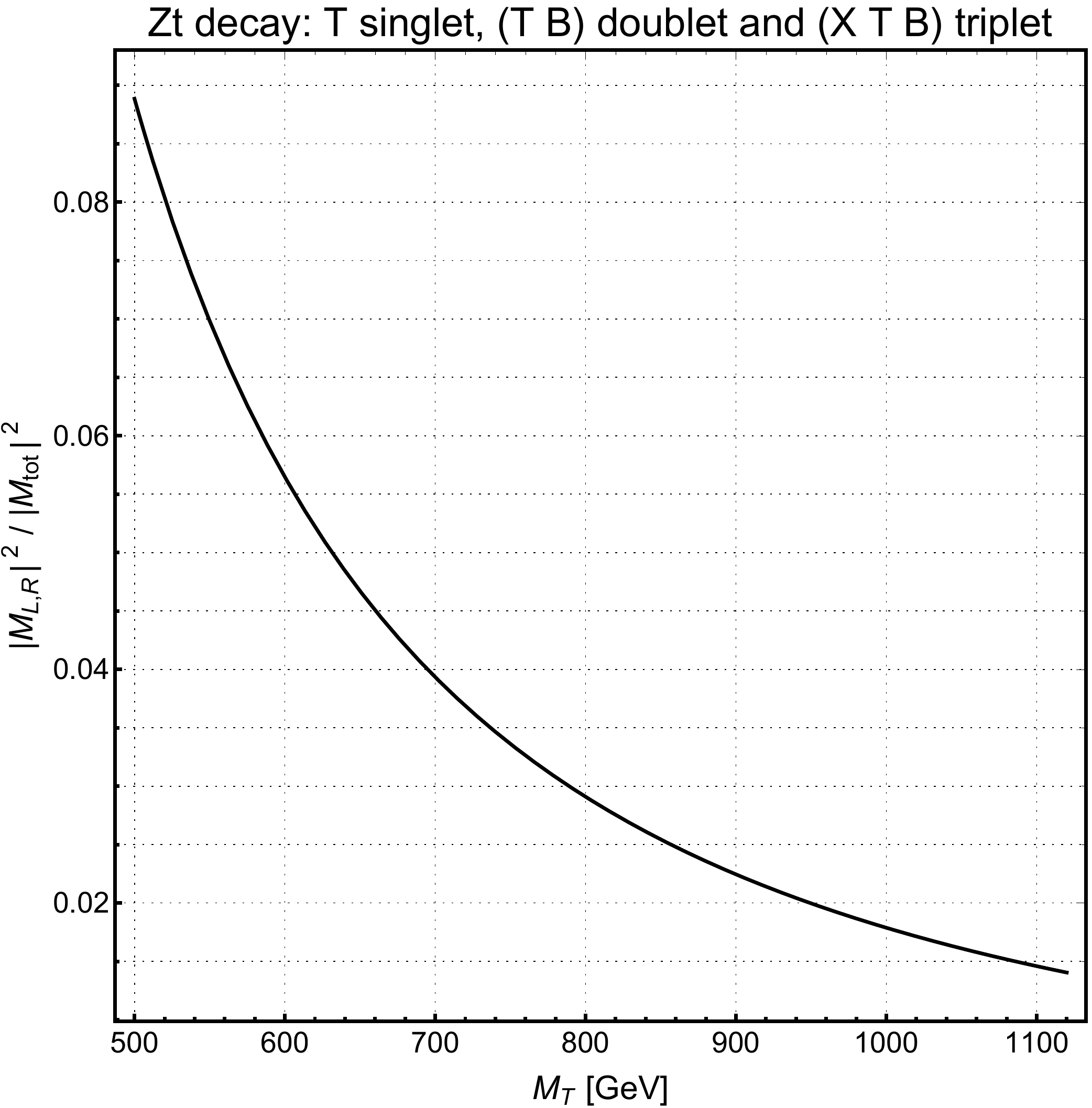}
\includegraphics[width=0.48\textwidth]{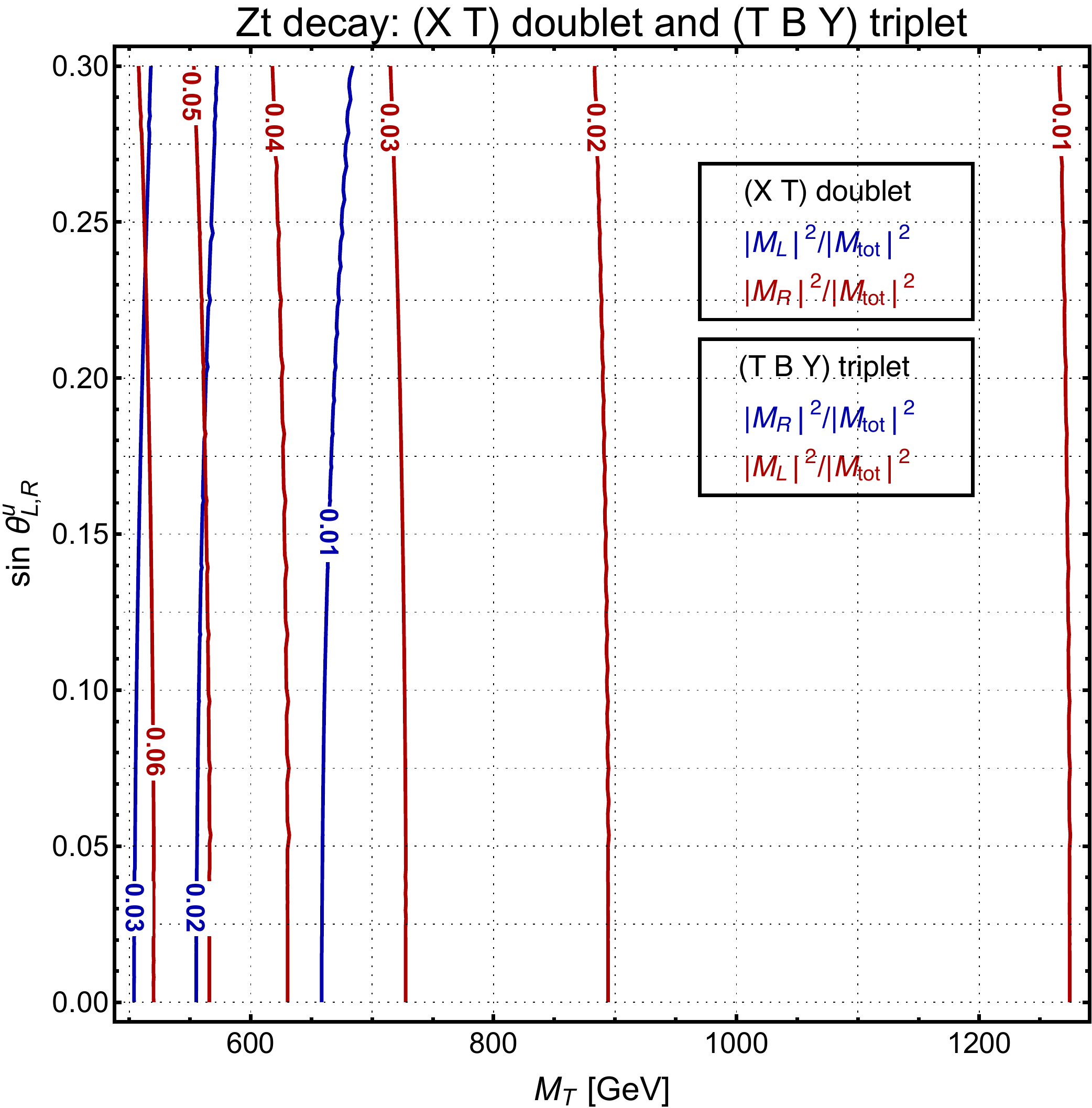}
\caption{Fraction of left-handed and right-handed polarised $Z$ boson arising from the neutral current decay of a $T$ VLQ. In the left panel the singlet, $(T~B)$ doublet and $(X~T~B)$ triplet representations are shown as function of $m_T$, while in the right panel the $(X~B)$ doublet and $(T~B~Y)$ triplet representations are shown in the $m_T-\sin\theta^u_{L,R}$ plane.}
\label{fig:zpol}
\end{figure}

For the neutral $Z$ boson  it is not possible to neglect the top mass. However the qualitative picture does not change significantly. The singlet and $(X~T~B)$ triplet do not possess a right-handed coupling and the same is true for the $(T~B)$ doublet and left-handed coupling. Thus, the ratio between the only non-zero transverse polarisation and the total contribution is independent of the mixing angle. Conversely, for the $(X~T)$ doublet and the $(T~B~Y)$ triplet there is a weak dependence on the mixing angle. We show in Fig.~\ref{fig:zpol} the same ratios of Fig.~\ref{fig:wpol} for all the possible $SU(2)_L$ representations of the  $T$ VLQ. 
Our results show that the $W$s and $Z$s bosons arising from the VLQ decay are always mainly longitudinally polarised. The longitudinal polarisation contribution increases with the mass of the VLQ and, for masses around the TeV, the transverse components fraction amounts to $\sim 1\%$.
This feature makes a measurement of the chirality of the VLQ coupling from gauge boson polarisations challenging, considering the uncertainties for the SM case reported in Tab.~\ref{tab:WpolarisationSMtop}.

\subsection{Top quark polarisation}
\label{sec:top-pol}

\begin{figure}[!htbp]
\centering
\includegraphics[width=0.48\textwidth]{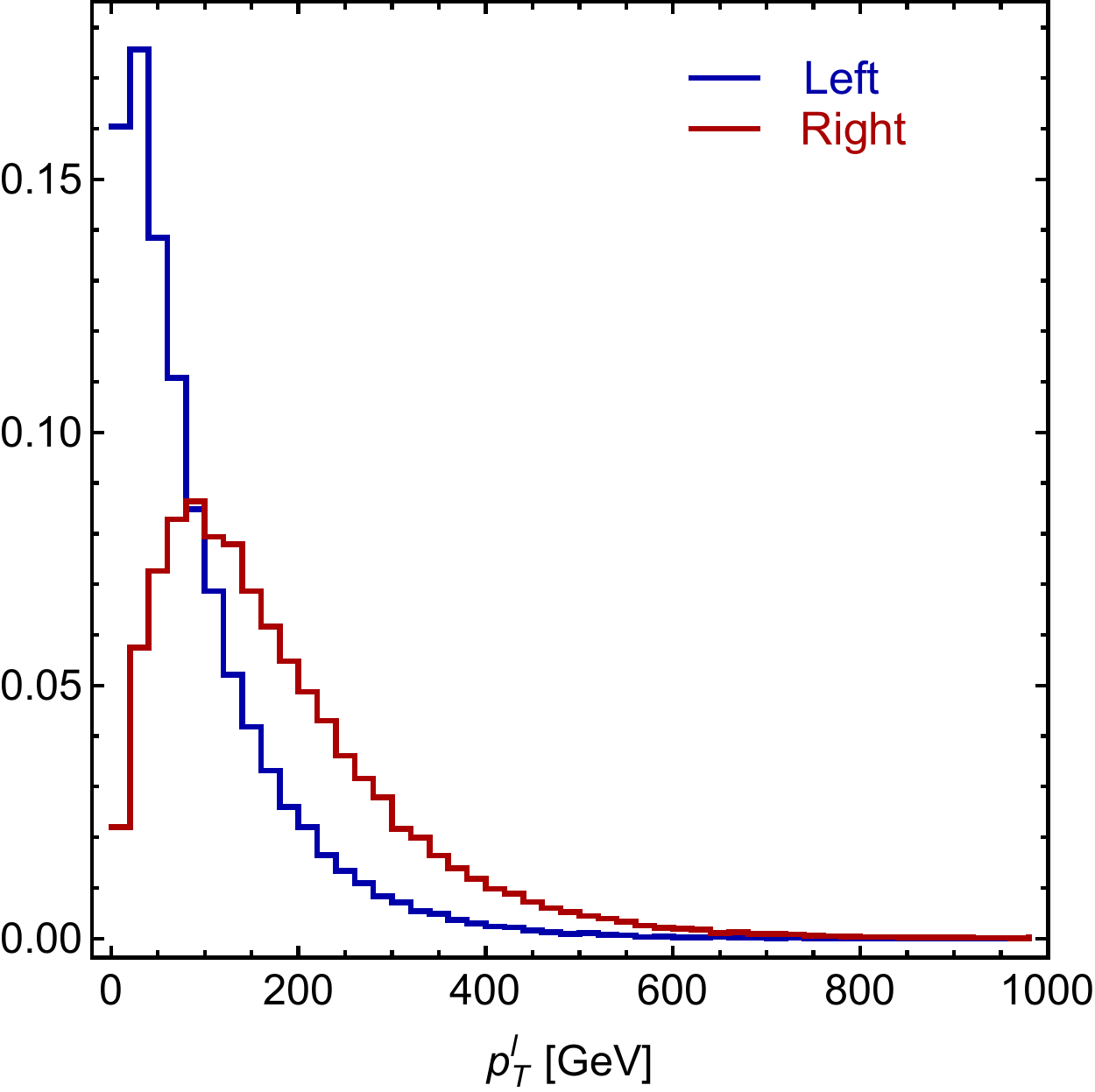}\hfill
\caption{\label{fig:LHC13-parton} Parton level shape of the transverse momentum distribution normalised to unit area of the lepton produced by the top quark arising from the decay of a pair-produced $X$ with $M_X=1$ TeV, at the 13 TeV LHC. The $X$ is decayed considering a purely left- or right-handed coupling with the top quark and $W$ boson.}
\end{figure}

Since the top quark decays before it hadronises, a second possibility can be offered by scrutinising its decay products, which can carry information on the polarisation of the top quark arising from the VLQ decay. The polar angle distribution of the top quark decay product $f$ in the top rest frame is described, see {\emph{e.g.}}~\cite{Belanger:2012tm}, by
\begin{equation}
\frac{1}{\Gamma_l} \frac{{\rm d} \Gamma_l}{{\rm d}\cos \theta_{f,{\rm rest}}}=\frac{1}{2}(1+\kappa_f P_t \cos\theta_{f,{\rm rest}})
\label{eq:pol-angle}
\end{equation}
where $\Gamma_l$ is the partial width, $\theta_{f,{\rm rest}}$ is the angle between the momentum of the decay product $f$ and the top spin vector, $\kappa_f$ is the analysing power of the decay product and $P_t$ is the polarisation of the top. In the case where the considered top decay product is a charged lepton one has $\kappa_f \sim 1$. From Eq.~\eqref{eq:pol-angle} one sees that for positive (negative) polarised top quarks most of the decay products come in the forward direction, that is the directions of the would-be momentum of the top quark in the laboratory frame. 
In the same frame the $\theta_{f}$ distribution is now described by Eq.~\eqref{eq:pol-angle} combined with a boost from the top rest frame to the laboratory frame. This implies that positive polarised top quarks will produce harder decay products. 
We illustrate this in Fig.~\ref{fig:LHC13-parton} where we show the parton level $p_T$ distribution normalised to unit area of the lepton produced by the top quark arising from the decay of pair produced $X$ VLQs, where the blue and red lines correspond to a purely left- and right-handed coupling of the $X$ VLQ with the SM top and $W$.
If a VLQ decaying to the SM top quark is observed at the LHC or future colliders, it is thus essential to provide a quantitative statement about the potential for discriminating the chirality of the VLQ coupling using information from the top polarisation. In the next Sections we will perform such analysis.

\section{Discovery and discrimination power of the LHC}
\label{sec:LHC}

In this Section we will study the prospects of the LHC in discriminating among the left and right-handed chiral coupling hypothesis focusing on the example of a $T$ VLQ decaying entirely into the SM top quark and $Z$ boson.

\subsection{Simulation details}

In order to study how the information from the top quarks arising from the VLQs decay can be used to disentangle the left- and right-handed coupling hypotheses and compare this results with the power the LHC has for discovering the VLQs, we recast a search for pair produced VLQs with charge 2/3 performed by the ATLAS collaborations in the single lepton channel with an integrated luminosity of 36.1 fb$^{-1}$ at $\sqrt{s}=13$ TeV~\cite{Aaboud:2017qpr}.
This search selects events with exactly one lepton, at least four jets and large missing transverse energy ($E_T^{\rm miss}$) and it has been designed to target VLQs with charge 2/3 decaying into a $Z$ boson and a top quark, setting a 95\% confidence level (CL) bound of 1160 GeV for a branching ratio (BR) of 100\% into the $Zt$ final state. For different BRs assumptions that correspond to the singlet representation and the doublet representation with $U(1)_Y$=1/6 the bounds read 870 GeV and 1050 GeV respectively. Note that in setting the bound the ATLAS collaboration neglected the kinematic difference between the singlet and the doublet case, always assuming the selection efficiencies derived for the $SU(2)_L$ singlet case, which thus produce a slightly conservative limit for the doublet case since, as shown in Fig.~\ref{fig:LHC13-parton}, a right-handed coupling structure gives rise to harder final state objects, which thus make easier for the events to pass the selections cuts, therefore increasing the signal selection efficiency\footnote{Other VLQs searches explicitly take into account this difference in selection efficiencies, which give rise to stronger limit for VLQ with a right-handed chiral coupling to SM bosons and quark, see {\emph{e.g}}~\cite{CMS:2017wwc}.}.

Our simulations have been performed using the VLQ model in {\tt UFO}~\cite{Degrande:2011ua} format of Refs.~\cite{FeynRulesVLQ,Buchkremer:2013bha} and used {\tt MadGraph5\_aMC@NLO}~\cite{Alwall:2014hca} as event generator. Parton showering, hadronisation and decay of unstable particles have been performed through {\tt PYTHIA\_v8}~\cite{Sjostrand:2014zea} while {\tt Delphes\_v3}~\cite{deFavereau:2013fsa} has been employed for a fast detector simulation. Jets have been reconstructed with {\tt FastJet}~\cite{Cacciari:2011ma}, via the {\tt anti-kT} algorithm~\cite{Cacciari:2008gp} with cone radius 0.4 using a tuned ATLAS detector card suitable for performing an analysis with {\tt MadAnalysis5}~\cite{Conte:2012fm}, which we have used as a framework to implement the ATLAS selection cuts. With this procedure and adopting the background information of the experimental analysis we found a 95\% CL mass bound for a $T$ VLQ which decays entirely into $Zt$ of 1143 and 1175 GeV for the left-handed and right-handed coupling case respectively.
Such bounds are within $\sim$1\% from the ones obtained by the ATLAS collaboration, thus validating our implementation.  These bounds have been derived estimating the statistical significance $\alpha$ as
\begin{equation}
\alpha = \frac{S}{\sqrt{S+B+ \epsilon_{\rm syst}^2 B^2}}
\end{equation}
where $S$ and $B$ stand for the number of signal and background events respectively and $\epsilon_{\rm syst}$ to the systematic uncertainty on the background determination, $\epsilon_{\rm syst}=\Delta B/B$. 

\subsection{Results}

\begin{figure}[!htbp]
\centering
\includegraphics[width=0.48\textwidth]{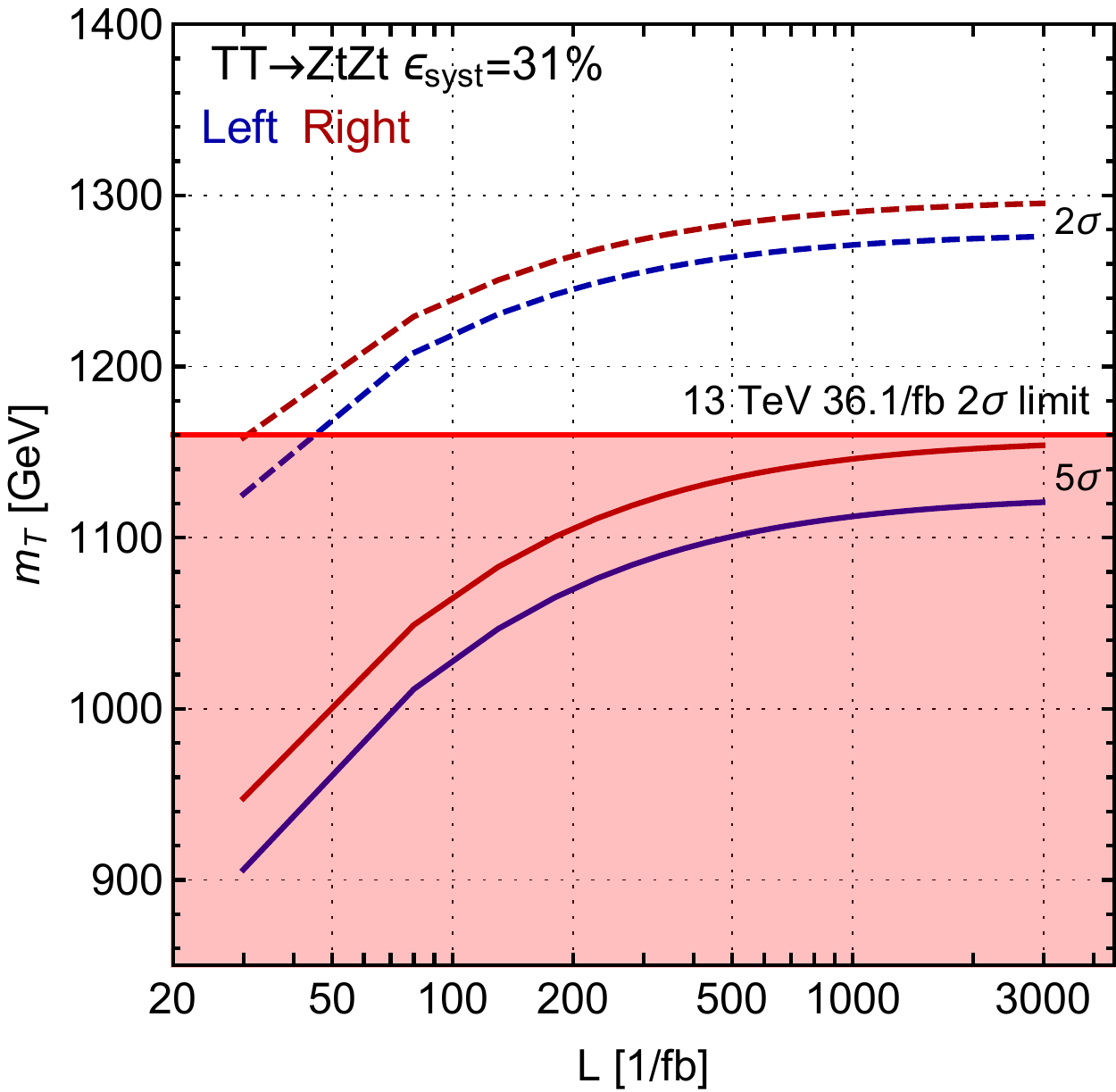}\hfill
\includegraphics[width=0.48\textwidth]{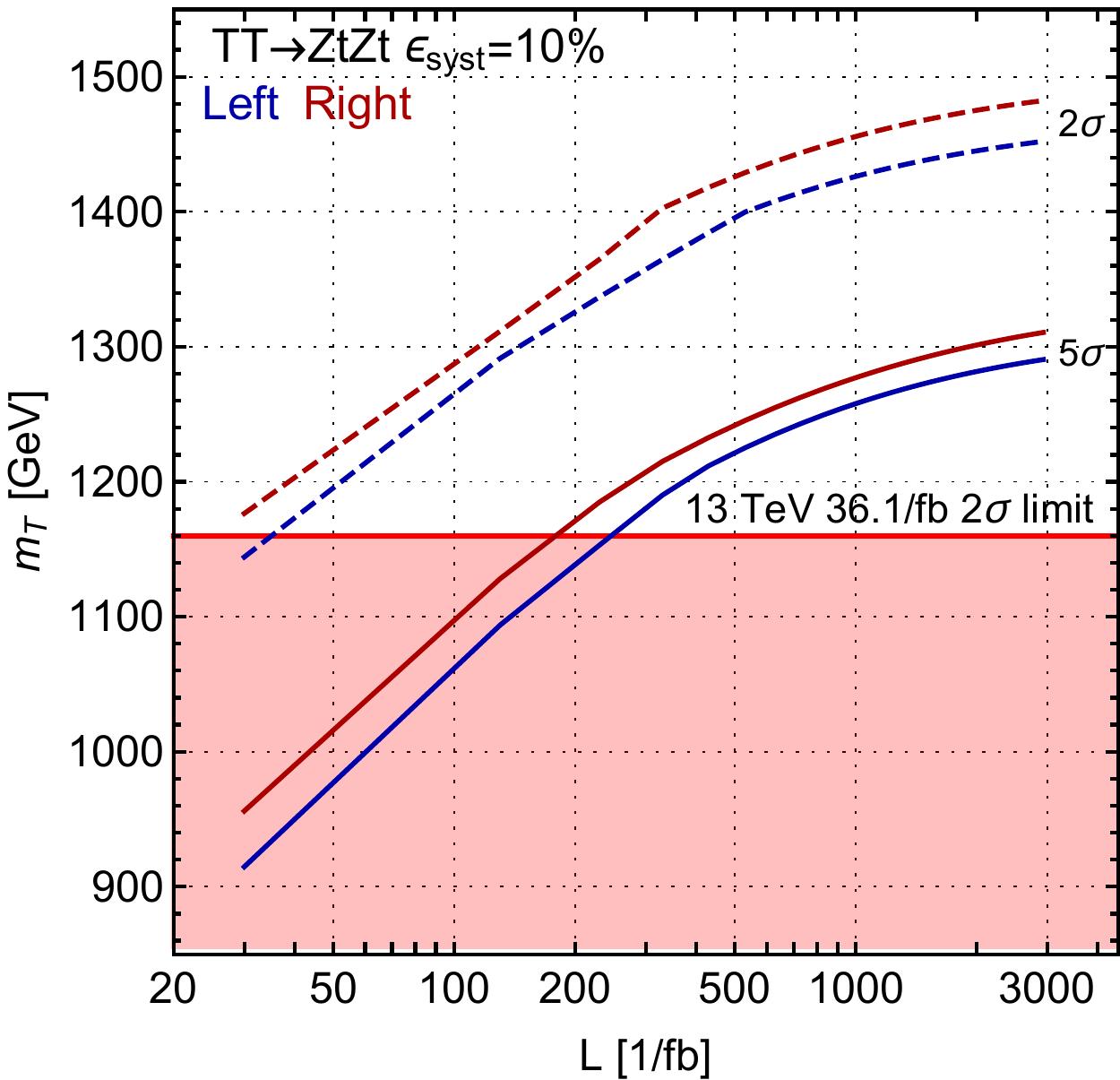}\hfill
\caption{\label{fig:atlas-proj}Projected exclusion (dashed) and discovery (solid) reach of the ATLAS single lepton search~\cite{Aaboud:2017qpr} for a $T$ VLQ decaying with 100\% branching ratio in $Zt$. The blue and red lines correspond to a left- and right-handed chiral coupling. The left and the right plots correspond to assuming an uncertainty on the background determination of 31\% and 10\% respectively. The red shaded area correspond to the present observed limit for a VLQ decaying with 100\% branching ratio into the $Zt$ final state, namely 1160 GeV~\cite{Aaboud:2017qpr}}
\end{figure}

The mass bounds for higher LHC integrated luminosities can be extrapolated by rescaling the numbers of signal and background events. The results are shown in Fig.~\ref{fig:atlas-proj} for both the left- and right-handed coupling structure as exclusions and discovery contours (corresponding to 2$\sigma$ and 5$\sigma$ respectively) for two choices of systematic uncertainties on the background determination: $\epsilon_{\rm syst}=31$\%, corresponding to the current systematic uncertainties of the ATLAS analysis~\cite{Aaboud:2017qpr}, and $\epsilon_{\rm syst}=10$\%.
With the selections and cuts of \cite{Aaboud:2017qpr}, and considering a maximum integrated luminosity of 3~ab$^{-1}$ it will be possible to exclude a VLQ with charge 2/3 decaying into $Zt$ up to $\sim 1300\;(1500)$ GeV with $ \epsilon_{\rm syst}=31$\% (10\%), while a discovery in this channel will only be possible if the uncertainty on the background is reduced from its current value: with $ \epsilon_{\rm syst}=31$\%, the discovery potential at 3~ab$^{-1}$ is already lower than the current exclusion limit, but it goes up to $\sim 1300$ GeV if $ \epsilon_{\rm syst}=10$\%. 

\begin{figure}[h!]
\centering
\includegraphics[width=0.48\textwidth]{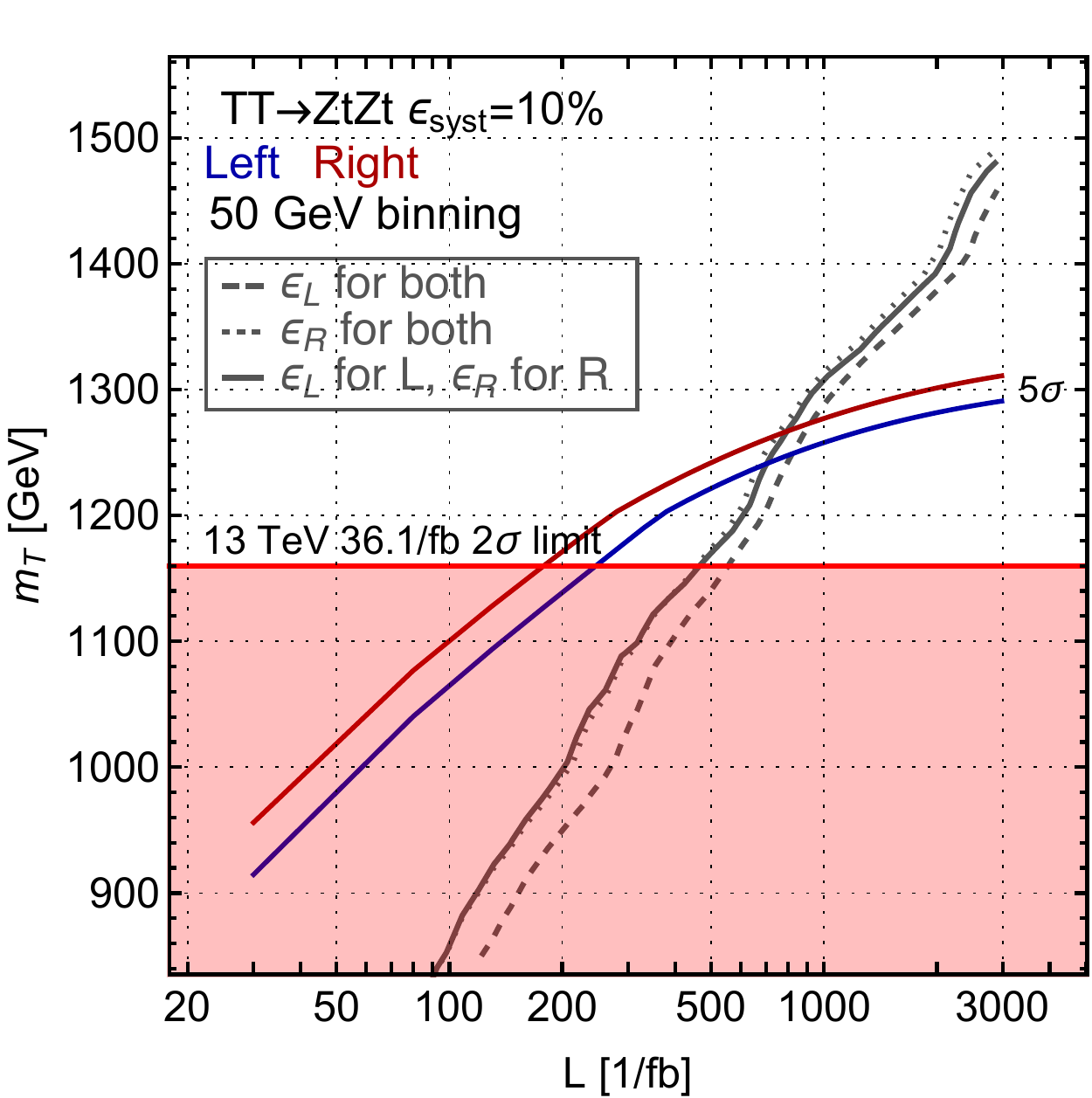}\hfill
\includegraphics[width=0.48\textwidth]{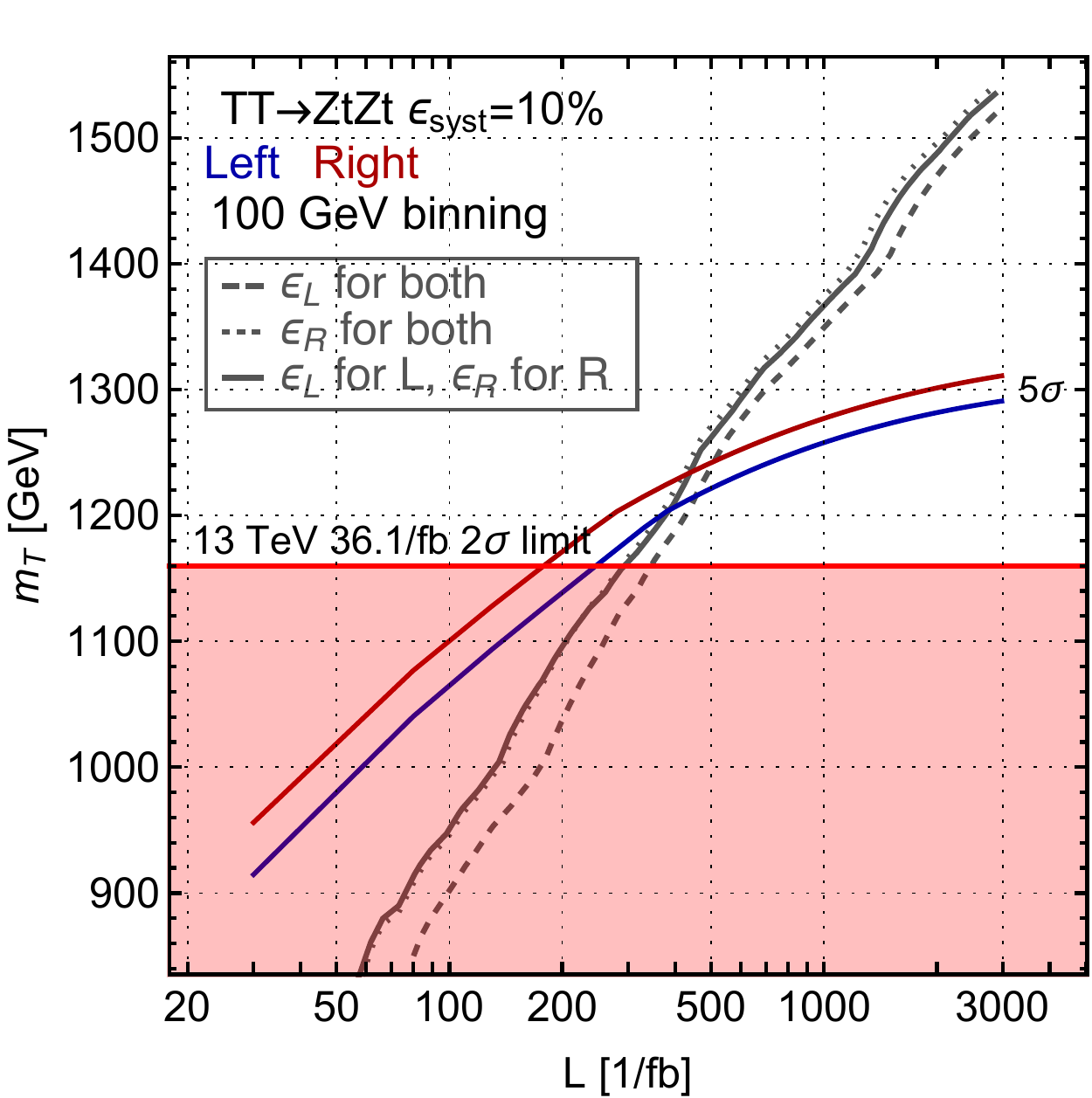}\\[10pt]
\includegraphics[width=0.48\textwidth]{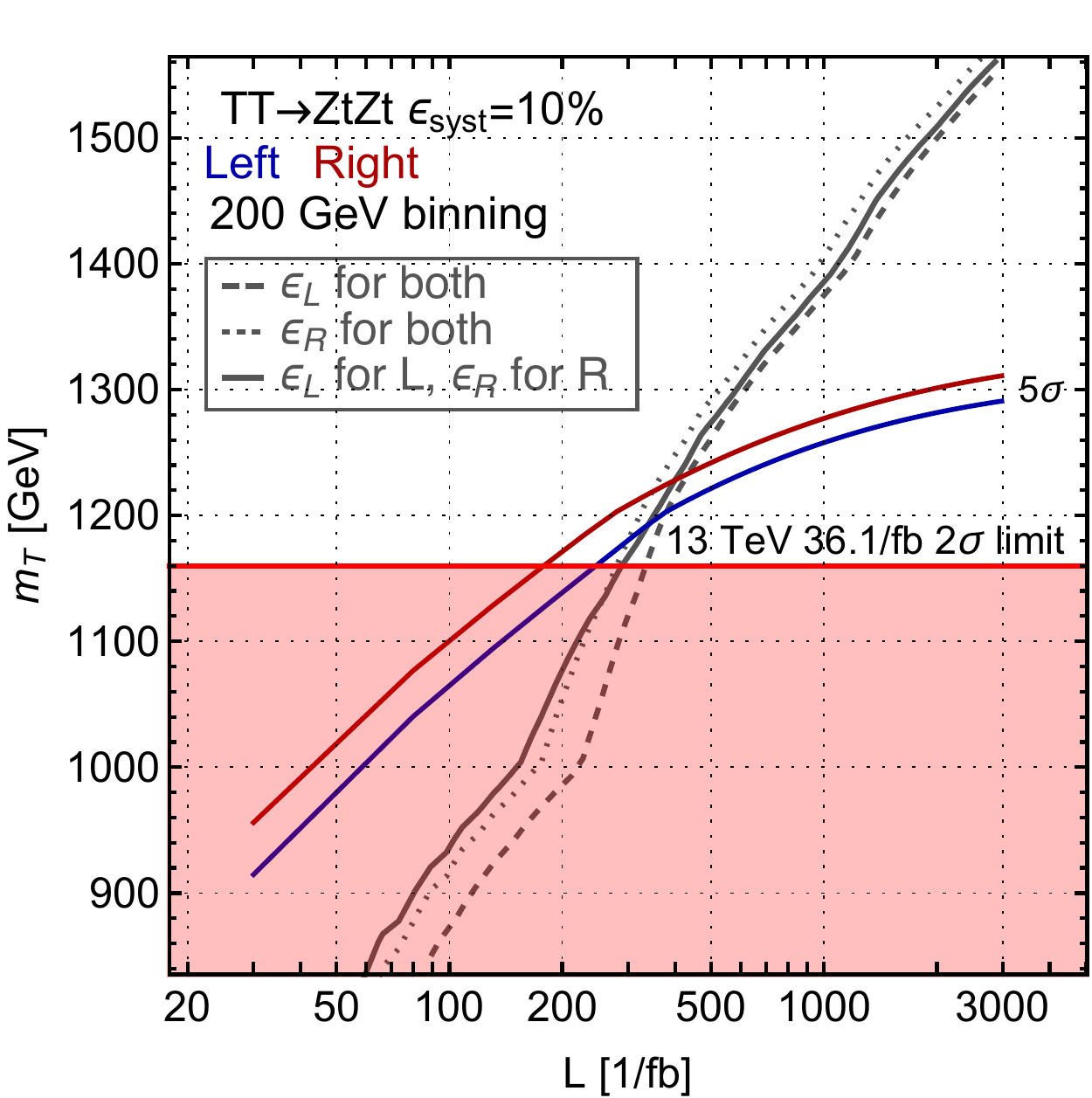}\hfill
\includegraphics[width=0.48\textwidth]{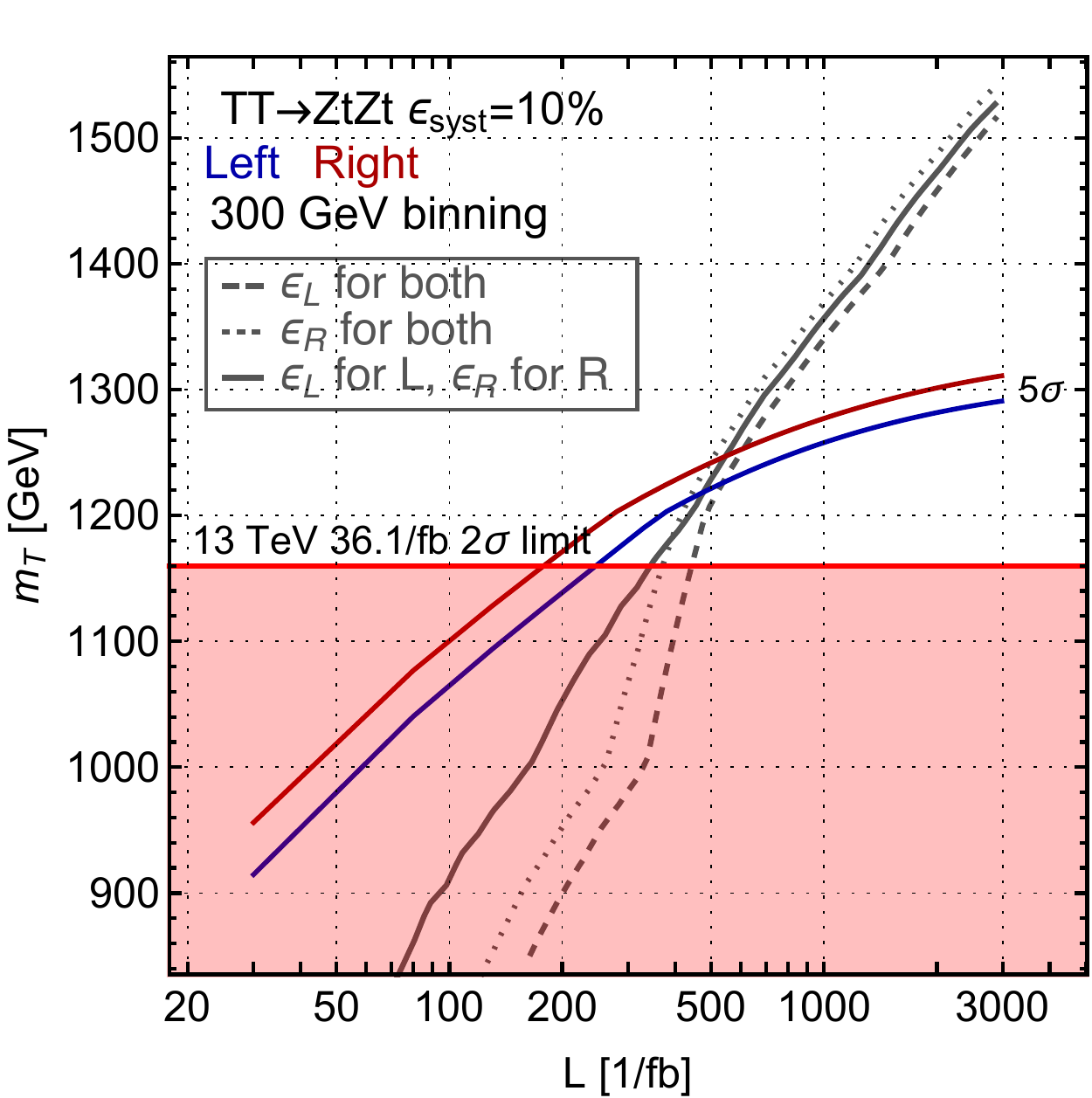}\\
\caption{\label{fig:13-discr}Projected 5$\sigma$ discovery (blue and red) and 2$\sigma$ discrimination (gray) reach of the ATLAS single lepton search~\cite{Aaboud:2017qpr} for a $T$ VLQ decaying with 100\% branching ratio into the $Zt$ final state. The blue and red lines correspond to the discovery reach for a the left-handed and right-handed coupling structure assuming an uncertainty on the background determination of 10\%. The red shaded area correspond to the present experimental limit from for a VLQ decaying with 100\% branching ratio into the $Zt$ final state, namely 1160 GeV~\cite{Aaboud:2017qpr}}
\end{figure}

Assuming then that the uncertainties on the background will be reduced to allow the possibility of discovering a $T$ VLQ in the $Zt$ channel, once the VLQ is discovered and its mass is determined with some degree of precision, it is in principle possible to exploit the differences between kinematic distributions in the left- and right-handed coupling hypotheses to determine the coupling structure of the VLQ. In the case under study, we will focus on the unique lepton present in the final state after the selection cuts have been imposed. Since the ATLAS analysis requires $E_T^{\rm miss}>300$ GeV, the selected lepton will come, most likely, from a leptonically decaying top quark, with the two $Z$ bosons decaying invisibly. The transverse momentum distributions of the lepton will thus be different for the left- and right-handed coupling hypotheses according to the top quark polarisation, as discussed in Sec.~\ref{sec:top-pol} .

We now assume that the signal and background distributions can be completely separated and perform a $\chi^2$ fit on the lepton $p_T$ distributions between the left- and right-handed hypotheses, with a number of degrees of freedom for the $\chi^2$ equal to the number of considered bins in the distributions:
\begin{equation}
\chi^2 = \sum _{i=1}^{n_{\text{bins}}} \frac{(L_i - R_i)^2}{\max \left(L_i,R_i \right)},
\end{equation}
where $L_i$ and $R_i$ are the events in the bin $i$ for left- and right-handed distributions; we have considered a Poissonian distribution of the events in the bins, such that their uncertainty is $\sqrt{N}$. The sum runs over all bins in which there are either $L$ or $R$ events, however due to the uniform binning there may be bins with no events in both distributions followed by bins with non-zero events, especially in the tails which are mostly affected by statistical noise; the number of degrees of freedom takes into account also such bins, and therefore our estimate will be conservative.
With this procedure it is possible to determine, for any VLQ mass, the integrated luminosity which is needed for disentangling the two hypotheses. The results are shown in Fig.~\ref{fig:13-discr}, where we compare the $5\sigma$ discovery reach, blue and red lines for the left- and right-handed coupling hypotheses respectively, with the 2$\sigma$ discrimination potential, gray lines, for different binnings of the lepton $p_T$ distribution. 
A comment here is in order. Beside the pure shape differences between the $p_T$ distributions for the two coupling hypotheses, also the total event normalisations for the two cases are not equal, thus potentially affecting the $\chi^2$ procedure. This is a consequence of the differences in selection efficiencies between the left- and right-handed coupling cases which however is not dramatically different, a fact reflected in the very similar 5$\sigma$ reach for left- and right-handed case. The discrimination power of our procedure is thus expected to be dominated by the pure difference in shape among the two distributions. In order to show this, we plot in Fig.~\ref{fig:13-discr} the 2$\sigma$ discrimination contours for three different choices of final events rate normalisation for the left- and right-handed coupling cases: normalising them to a common value, respectively either the one corresponding to the left- or right-handed hypothesis, and normalising each distribution to its true value, {\emph{i.e.}} taking into account the differences in selection efficiencies.
As expected the qualitative picture does not change significantly. The three discrimination curves intersect the 5$\sigma$ discovery contour at roughly the same point.
 The binning of the $p_T$ distributions has a stronger impact: a discrimination can be obtained preferably with a binning around 100-200 GeV, while it becomes less performant for smaller or larger binning. Of course, a proper optimisation of the bin size, possibly considering non-uniform bins, would be in order, but this goes beyond the purposes of this explorative analysis.

Numerically, we obtain that if a VLQ with a mass lighter than 1200 GeV is discovered, a mild increase in integrated luminosity will be needed to disentangle the two hypotheses, while the collected dataset will already be enough for the discrimination if a VLQ heavier than 1200 GeV is found with a $5\sigma$ significance.

\section{Discovery and discrimination power at 33 and 100 TeV hadron colliders}
\label{sec:FCC}

The results of Section~\ref{sec:LHC} show that with the signal region currently used in the ATLAS search we have considered, the LHC discovery reach will mildly increase when further data will be collected. As shown in Figs.~\ref{fig:atlas-proj} and \ref{fig:13-discr}, passing from 100 to 3000 fb$^{-1}$ of integrated luminosity the 5$\sigma$ mass reach will only increase from 1100 to 1300 GeV, for an optimistic assumption on the future determination of the background uncertainty. 
Analogous results are reported in different analyses: in Ref.~\cite{CMS:2013xfa} for example, a discovery reach of $\sim950$ GeV and 1300 GeV is estimated for the dilepton and single-lepton channel at the 14 TeV LHC with 300 fb$^{-1}$ of integrated luminosity. These values increase respectively to 1150 and 1500 GeV with 3 ab$^{-1}$ collected; 
in Ref.~\cite{Matsedonskyi:2014mna} the exclusion reach for a pair-produced $X_{5/3}$ decaying to $Wt$ at 13 TeV is estimated to be around 1400 and 1700 GeV with 100 fb$^{-1}$ and 3 ab$^{-1}$ of integrated luminosity respectively.
In general, the reach at a hadron colliders is limited, for fixed energy, by the fact that the contribution of the PDFs drops when the transferred momentum of the process approaches the kinematic limit $\sqrt{s}/2$.

\begin{figure}[htbp!]
\centering
\includegraphics[width=0.48\textwidth]{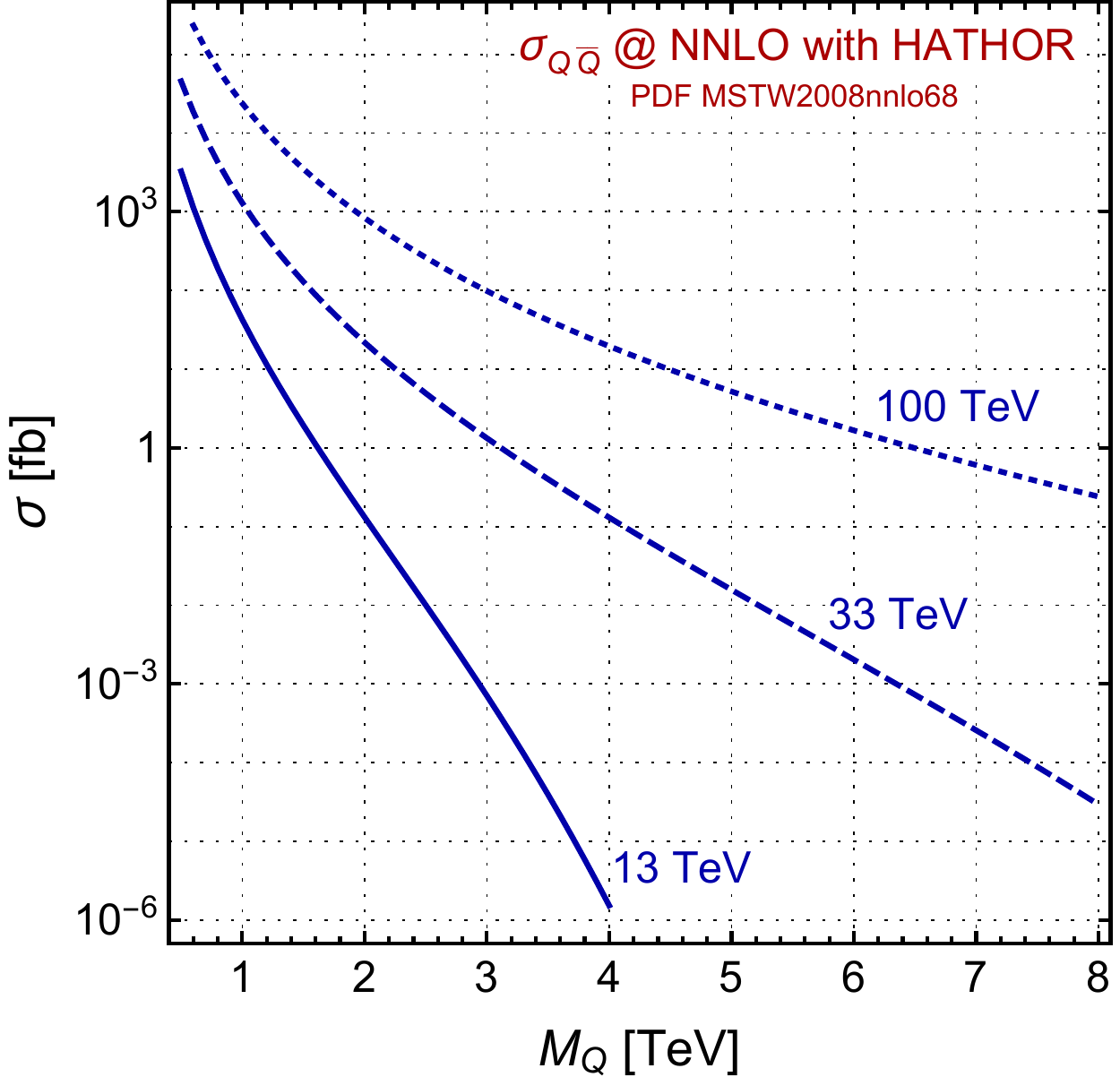}
\caption{\label{fig:sigmas} Cross-sections for QCD induced heavy quark pair production at proton-proton colliders with different centre of mass energies.}
\end{figure}

If VLQs are heavier than $\sim$1.5 TeV, thus eluding the projected 2$\sigma$ bounds of current analyses, and with the LHC about to enter its Run III phase in a few years, it is crucial to evaluate the reach of the proposed next generation high-energy hadron colliders for discovering (or excluding) heavier BSM physics. At the same time, it is also important to evaluate the potentialities of such prototypes to discriminate between different theories, should NP be found during their operations. It is thus the purpose of this Section to give an estimate of what is the mass reach of future high energy hadron colliders for excluding or discovering pair-produced VLQs and how these projected bounds compare with the discrimination power of such machines. 
We will focus on proton-proton colliders with a centre of mass energy of 33 and 100 TeV which are standard benchmark energies for proposed future hadron colliders. The quark pair-production cross-sections for these centre of mass energies, computed at NNLO using {\tt HATHOR}~\cite{Aliev:2010zk} with {\tt MSTW2008nnlo68} PDFs~\cite{Martin:2009bu}, are illustrated in Fig.~\ref{fig:sigmas}. As illustrative example we will present our results for the representative case of an exotic VLQ with charge 5/3, again coupled only to third generation quarks. 
This state interacts only through charged current and undergoes the decay $X\to W^+ t \to W^+ W^+ b$, which can give rise to a same-sign dilepton (2SSL) final state, a commonly used search channel for the $X$ state~\cite{Chatrchyan:2013wfa,Sirunyan:2017jin}.
For our analysis, we closely follow the search strategy defined in~\cite{Avetisyan:2013rca} where the authors have designed an analysis for the search of the $X$ VLQ in the 2SSL channel at the 14 TeV LHC with high luminosity option, as well as at a future 33 TeV proton proton collider. We thus extrapolate their results for the case of a 100 TeV collider.

\subsection{33 TeV}

The search strategy proposed in~\cite{Avetisyan:2013rca} requires the presence of two same-sign leptons with a transverse momentum greater than $150$ and $50$ GeV respectively, as well as at least two jets with the same $p_T$ thresholds. Events are further selected by asking $E_T^{\rm miss}>200\;$GeV, $H_T>2200\;$GeV and $S_T>3000\;$ GeV, where $H_T$ is the scalar sum of the transverse momentum of all jets and leptons in the events and $S_T=H_T+E_T^{\rm miss}$. Additionally, we impose a $Z$ boson veto for the di-electron channel and, for tri-lepton final states, between any of the two selected leptons and any other same flavour and opposite sign lepton. Both vetoes are in the mass range 76-106 GeV. Finally, events should contain at least seven objects, including the two selected leptons. A comment here is in order. The analysis of~\cite{Avetisyan:2013rca} makes use of top-tag and $W$-tag algorithms, with a top-tag jet and a $W$-tag jet counting as three and two constituents respectively. In our analysis however we did not implemented these  boosted object reconstructions procedures, only relying on a standard anti-$k_T$ clustering algorithm with a 0.5 cone radius for signal jets. As we will show however, the signal selections efficiencies that we obtained are in good agreement with the one obtained in~\cite{Avetisyan:2013rca} thus validating a posteriori our results. Differently from the authors of~\cite{Avetisyan:2013rca}, we also did not take into account pile up effects throughout our simulation. We implemented the selections cuts described above in the {\tt MadAnalysis5} framework and simulated our signal samples for a $m_X$ mass between 2 and 3 TeV following the same procedure described in Sec.~\ref{sec:LHC} for the 13 TeV case. The selection efficiencies we obtain are reported in Tab.~\ref{tab:33eff} and, albeit slightly higher, they are overall in good agreement for the left-handed coupling case.

\begin{figure}[!htbp]
\centering
\includegraphics[width=0.48\textwidth]{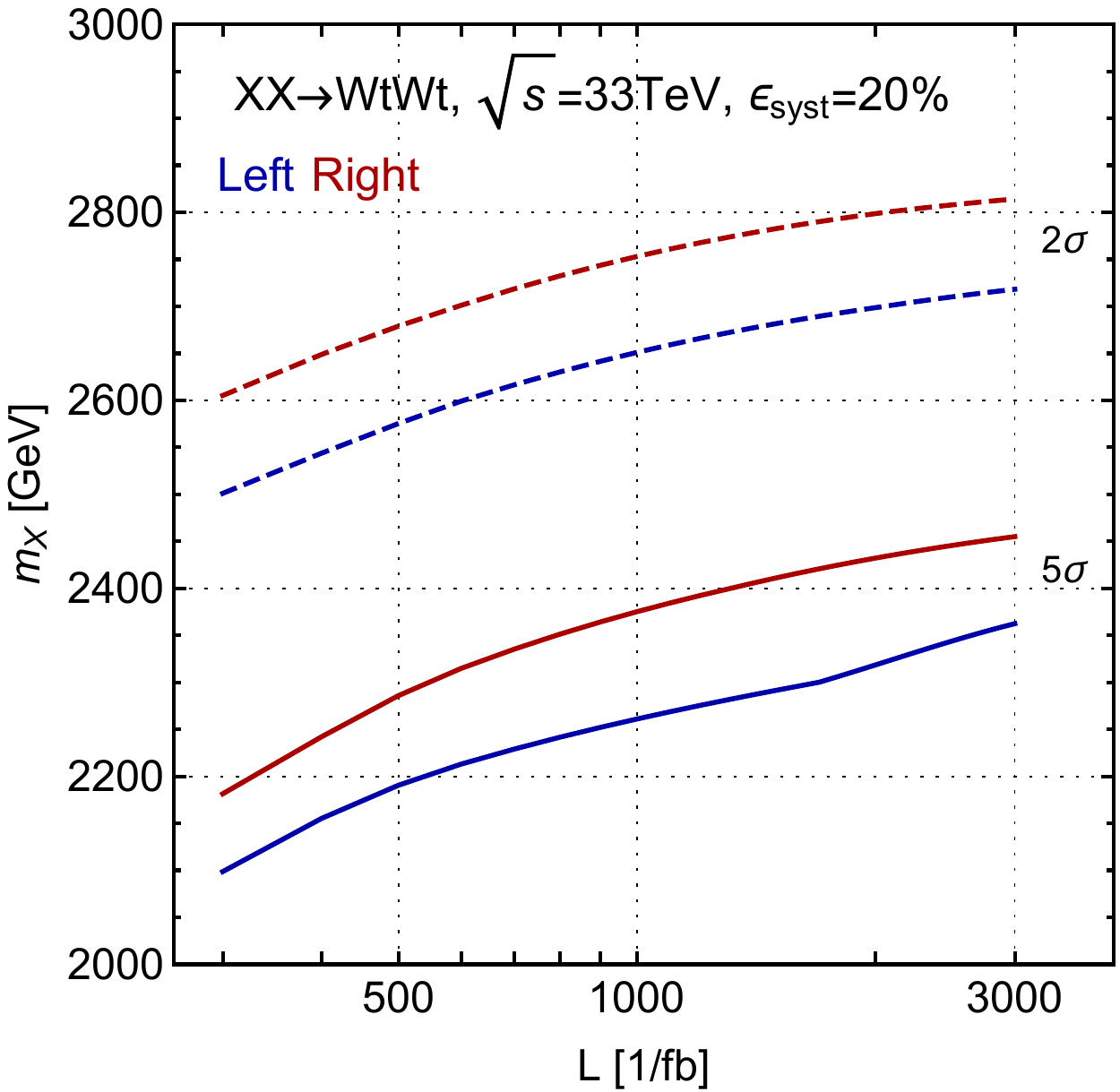}
\caption{\label{fig:33Lumi_1} Projected exclusion (dashed) and discovery (solid) reach of the same-sign dilepton search for a $X$ VLQ decaying with 100\% branching ratio into $Wt$ for $\sqrt{s}=33$ TeV hadron collider. The blue and red lines correspond to a left- and right-handed chiral coupling of the $X$ VLQ. We assume an uncertainty on the background determination $\epsilon_{\rm syst}=20\%$.}
\end{figure}

\begin{table}[htbp]
\centering
\begin{tabular}{c|c|c}
\toprule
$m_X$ [TeV] 		& Our eff. (L/R) [$\times 10^{-3}$]	&	Eff. [$\times 10^{-3}$]~\cite{Avetisyan:2013rca} \\
\midrule
\midrule
	2					&		7.9\;/\;11.1					&	6.4			\\
	2.1				& 			8.6\;/\;11.1					& 	7.0  \\
	2.2				& 			8.6\;/\;11.1					& 	6.4\\
	2.3				& 			7.9\;/\;11.5					& 	7.6\\
	2.4				& 			9.0\;/\;11.2					& 	7.1\\
	2.5				& 			8.5\;/\;11.2					& 	6.9\\
	2.6				& 			8.6\;/\;11.3					& 	7.4\\
	2.7				& 			8.7\;/\;11.3					& 	7.0\\
	2.8				& 			8.9\;/\;11.3					& 	6.8\\
	2.9				&	 		8.3\;/\;10.9					&	6.3 \\
	3.0				&	 		7.8\;/\;10.9					&	6.9 \\
\bottomrule
\end{tabular}
\caption{\label{tab:33eff}Selection efficiencies obtained through our simulation for the signal selection proposed in ~\cite{Avetisyan:2013rca} for a pair produced $X$ VLQ decaying into the $Wt$ final state compared with the one of~\cite{Avetisyan:2013rca}.}
\end{table}
In order to set the 95\% CL limit on the $X$ mass we use the background information with 50 mean pileup interactions per bunch crossing provided in~\cite{Avetisyan:2013rca}, and assume an uncertainty on the background determination of 20\%. We obtain a 2$\sigma$ bound of 2501 (2605) and 2718 (2814) GeV for the left (right) handed coupling hypothesis with 300 and 3000 fb$^{-1}$ of integrated luminosity, to be compared with the original results of 2480 and 2770 GeV respectively. These results are showed in Fig.~\ref{fig:33Lumi_1} where 2$\sigma$ and 5$\sigma$ contours are plotted for an integrated luminosity in the range 300-3000~fb$^{-1}$.

\begin{figure}[!htbp]
\centering
\includegraphics[width=0.48\textwidth]{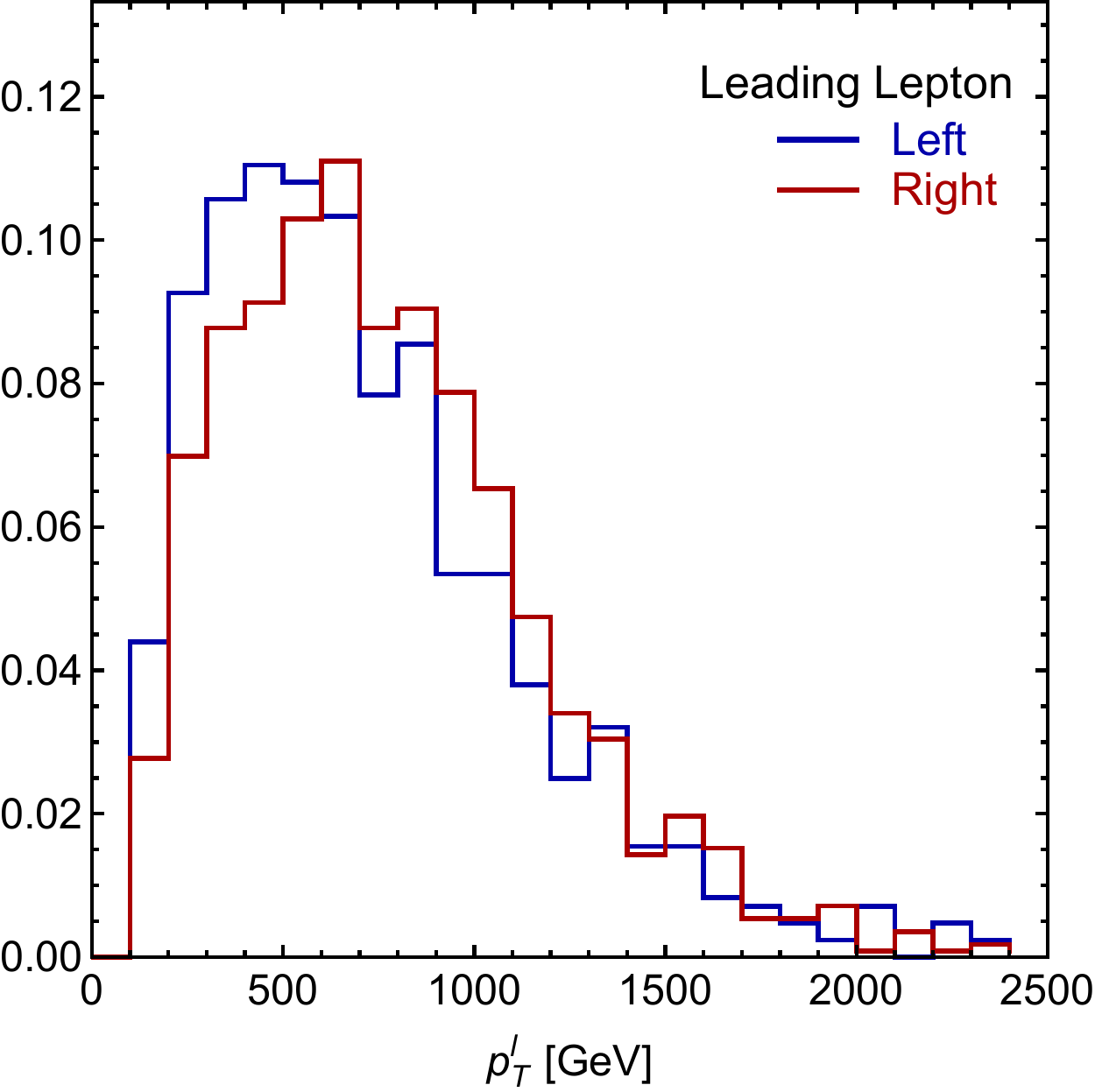}\hfill
\includegraphics[width=0.48\textwidth]{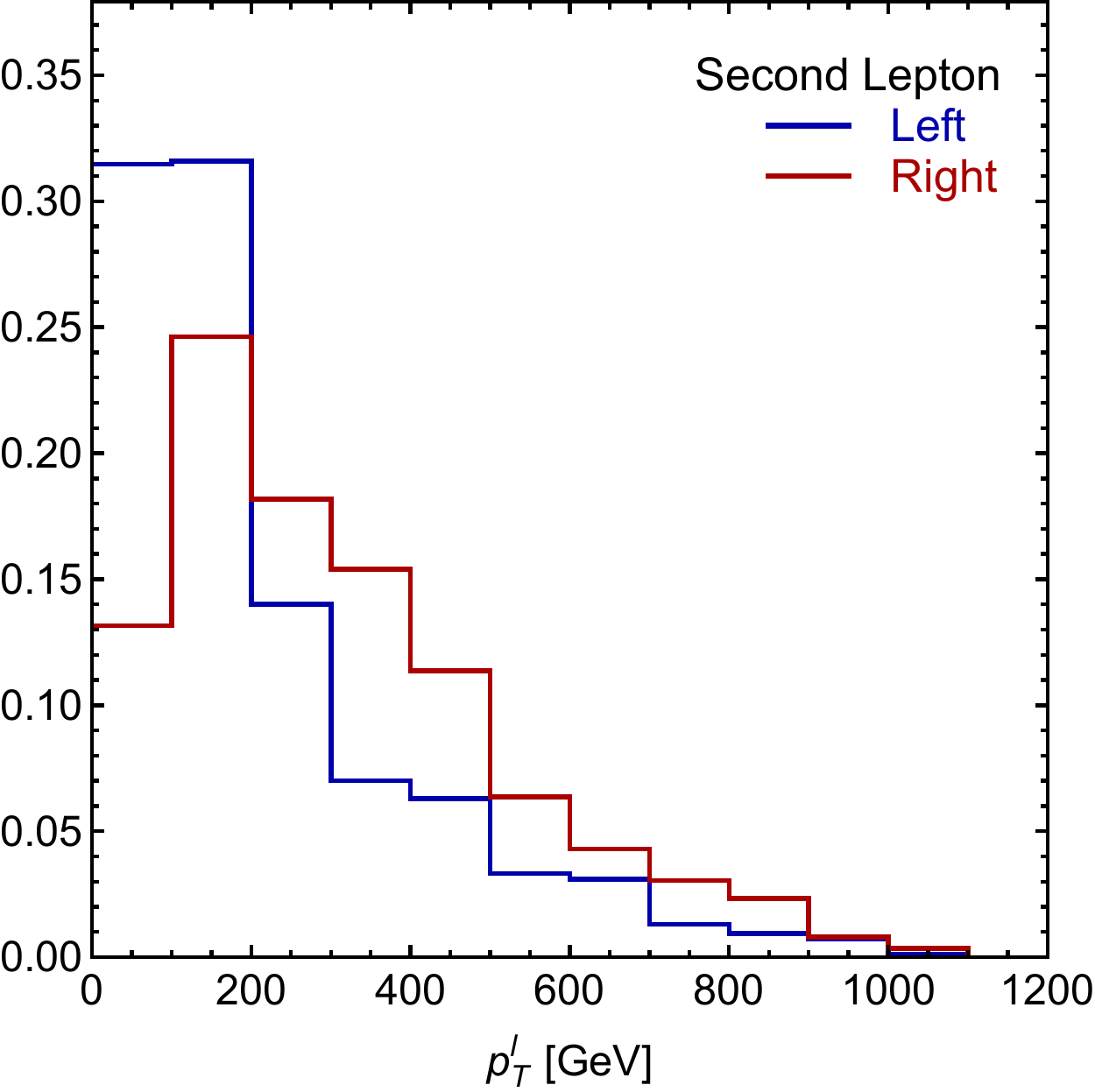}\hfill    
\caption{\label{fig:LHC33-detector} Detector level shape of the transverse momentum distribution normalised to unit area of the leading (left panel) and subleading (right panel) lepton in the 2SSL channel for left-handed and right-handed $X$ decaying into $Wt$ with $m_{X}=2500$ GeV and $\sqrt{s}=33$ TeV.}
\end{figure}

We want now to determine which observables, among those which can be built out of the 2SSL final state, can be exploited to discriminate among the two coupling hypotheses of the VLQ. Following the discussion of Sec.~\ref{sec:param} and the results for the 13 TeV LHC case, we will consider the $p_T$ spectrum of the two leptons in the selected events. An important difference with respect to the single lepton search considered in Sec.~\ref{sec:LHC} is that in this case the leptons might or might not arise from the decay of the top quark coming from the $X$ decay. However, it is not necessary to reconstruct the top quark system from its consitutents to select the correct leptons; it is enough to rely on the events sample surviving the selections cuts described above.
We illustrate this in Fig.~\ref{fig:LHC33-detector} where we show the detector level $p_T$ distributions of the hardest and second hardest lepton in the event, for $m_{X}=2500$ GeV. As it can be seen, the distribution of the second hardest lepton exhibits a larger difference between the left- and right-handed coupling hypotheses with respect to the leading lepton one. 
\begin{figure}[!htbp]
\centering
\includegraphics[width=0.48\textwidth]{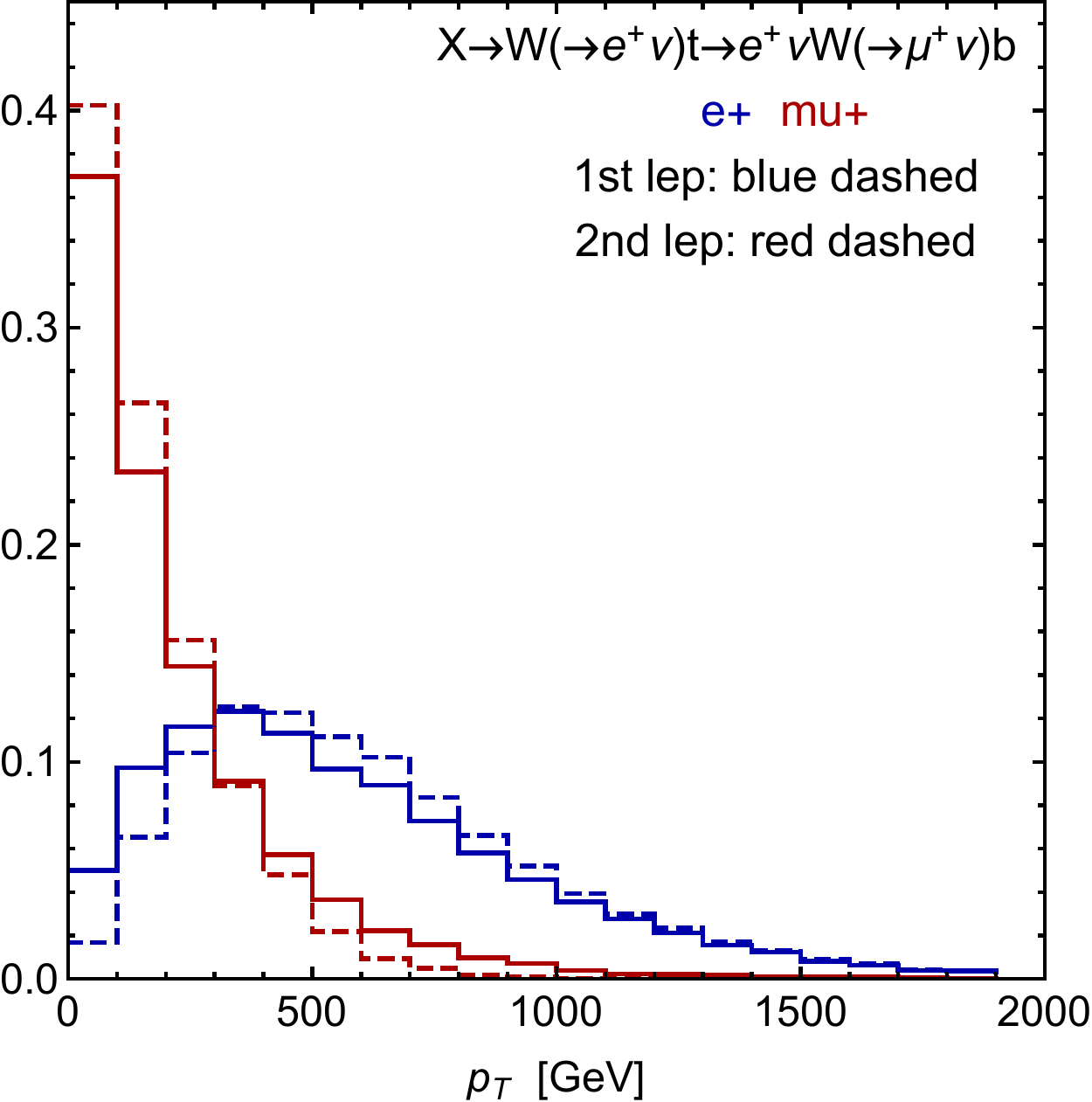}\hfill
\caption{\label{fig:lep_identification_33} Parton level shape of the transverse momentum distribution normalised to unit area of the electron and muon compared to the ones of the leading and subleading leptons in the process $p p \to X \bar X_{}$ with $X\to W^+(\to e^+ \nu_e) t$ and $t \to W^+(\to \mu^+ \nu_\mu) b$, with $m_{X}=2500$ GeV and $\sqrt{s}=33$ TeV.}
\end{figure}
This can be understood from the fact that in the decay process $X\to W^+ t \to W^+ W^+ b$ the lepton from the first $W^+$ boson carries almost always the highest transverse momentum, while the lepton arising from the top quark decay, which can be exploited for discrimination, is generically softer. To numerically quantify the previous statement we have simulated the process $p p \to  \bar{X}  X\to W^+ t \bar{X}$, forcing the decay of the first $W$ boson to the $e^+ \nu_e$ final state, while the $W$ arising from the top quark decay has been decayed to $\mu^+ \nu_\mu$. In Fig.~\ref{fig:lep_identification_33} the parton level transverse momentum of the electron, muon, leading lepton and second leading lepton are shown: the identification of leading and subleading leptons with the leptons arising from $W$s coming from $X$ and top is indeed very accurate. A further consequence of this kinematic feature is that our results for the $X$ VLQ can be easily interpreted also in terms of a charge -1/3 $B$ VLQ, considering the 2SSL channel. For pair-produced $B$ VLQs, the same-sign leptons will belong to opposite branches; however, the leading lepton will still originate from the $W$ coming directly from the $B$ and the sub-leading lepton will still arise from the decay of the top, the only difference being that the top is produced by the other $B$ quark.

\begin{figure}[!htbp]
\centering
\includegraphics[width=0.48\textwidth]{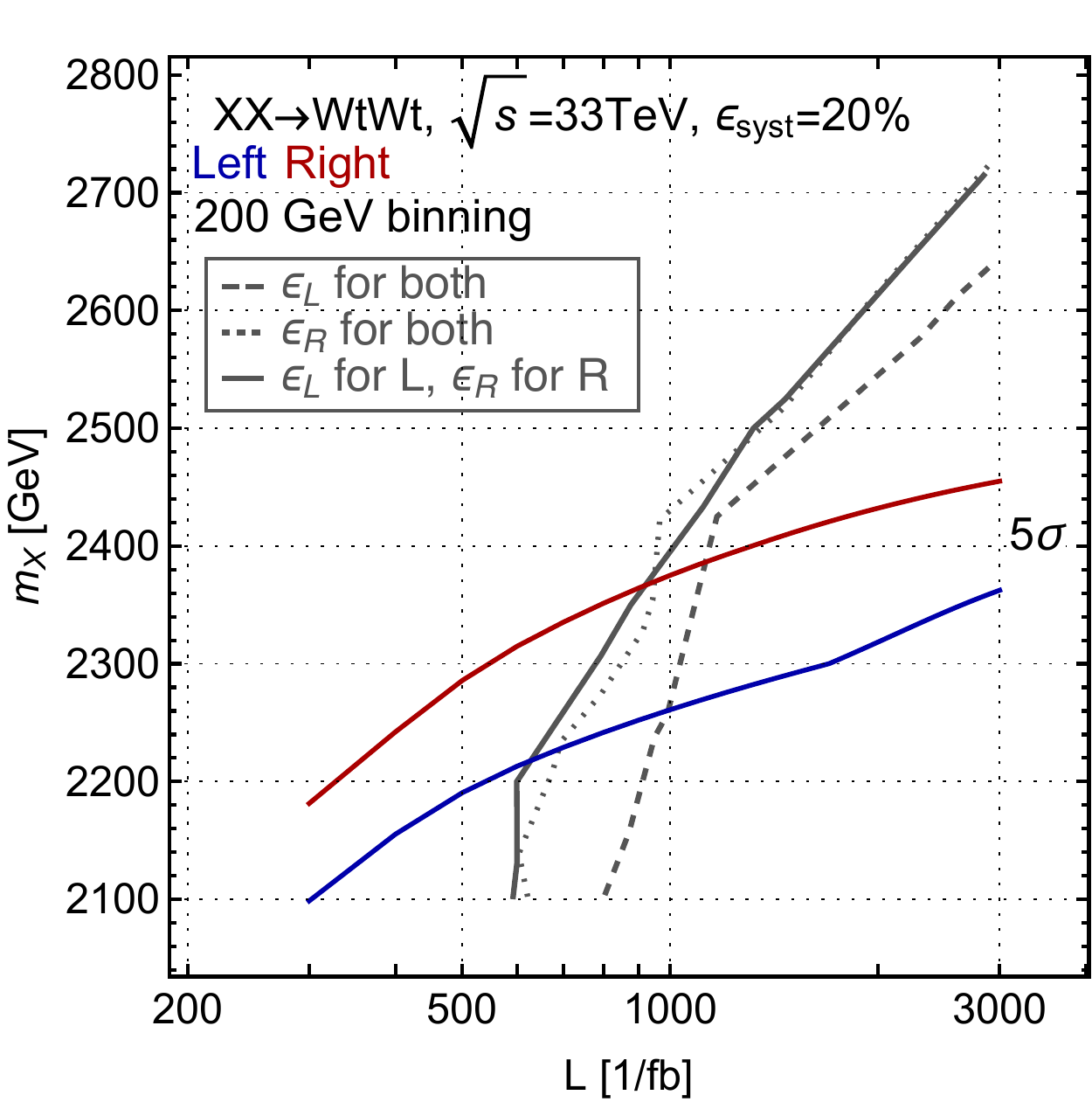}
\caption{\label{fig:33Lumi_2} Projected discovery (blue and red) and discrimination (gray)  reach of the 2SSL search for a $X$ VLQ decaying with 100\% branching ratio into the $Wt$ final state at $\sqrt{s}=33$ TeV. We use the $p_T$ distribution of the sub-leading lepton with a binning of 200 GeV and assume an uncertainty on the background determination of 20\%.}
\end{figure}

We therefore choose to use the distributions of the $p_T$ of the second hardest lepton in the events and perform a $\chi^2$ fit, analogously to what we did in Sec.~\ref{sec:LHC} for the 13 TeV LHC case. Our results are illustrated in Fig.~\ref{fig:33Lumi_2} where the gray lines represent the 2$\sigma$ discrimination contour between the left- and right-handed hypothesis with a binning of 200 GeV (we have verified that this binning choice optimises the discrimination power) and for different assignments of selection efficiencies for the two cases, as discussed for the 13 TeV analysis. Our results show again that a discrimination among the left- and right-handed coupling hypotheses is possible in all the discovery range accessible at a 33 TeV collider. In particular, if a VLQ with mass greater than $\sim 2300\;$ GeV is discovered, the collected data set will already be sufficient to exclude one of the two coupling structure.

\subsection{100 TeV}

To provide estimates for the discovery and discrimination reach of a 100 TeV hadron collider we consider the same process of a pair produced $X$ VLQ which decays into $Wt$ studied for the 33 TeV case, and we will analyse again the 2SSL channel. 
In doing so we we will adopt the following simplifying assumptions. We assume that the signal selection efficiencies on the $W^+W^- t \bar t$ final state remain constant to the ones obtained for the 33 TeV case and that do not depend on the mass of the $X$ VLQ. Furthermore we assume them to be equal for the left- and right-handed coupling cases. This efficiency has then been fixed to the value obtained in~\cite{Avetisyan:2013rca} for $m_X=3000\;$ GeV which is around $6.9 \times 10^{-3}$.
\begin{figure}[!htbp]
\centering
\includegraphics[width=0.48\textwidth]{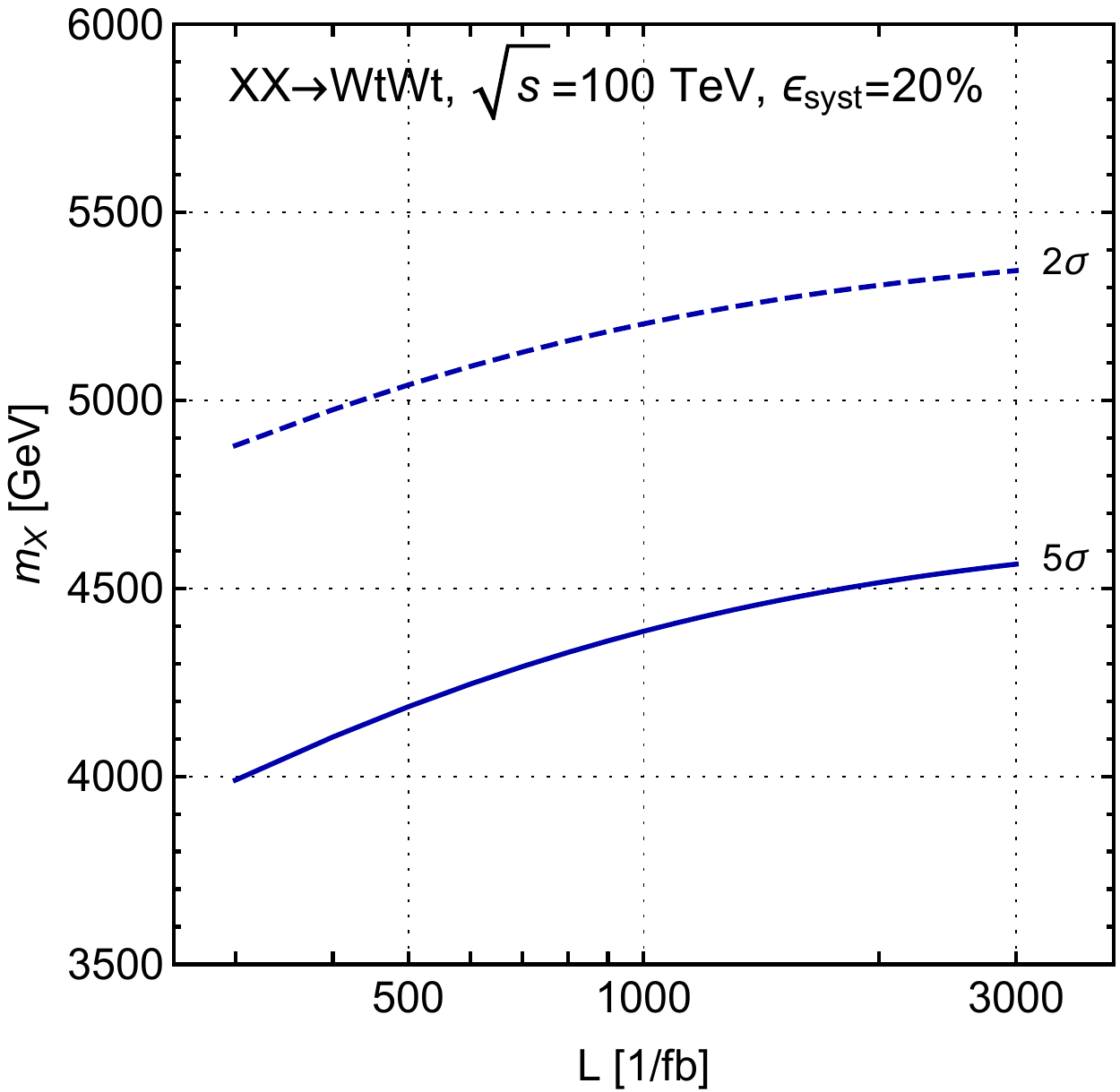}\hfill
\includegraphics[width=0.48\textwidth]{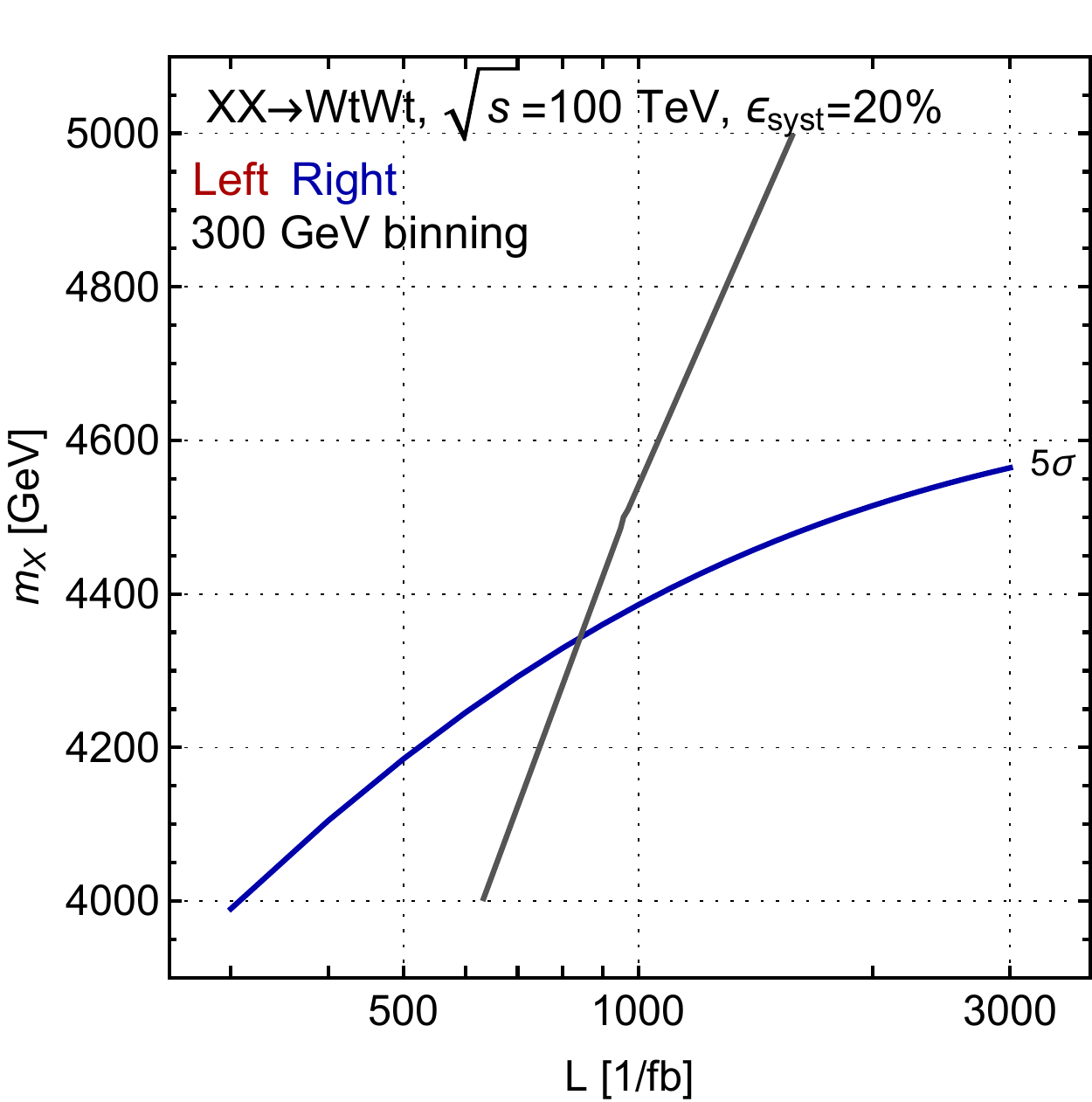}
\caption{\label{fig:100Lumi}Left panel: Projected exclusion (dashed) and discovery (solid) reach of the same-sign dilepton search for a $X$ VLQ decaying with 100\% branching ratio into $Wt$ for $\sqrt{s}=100$ TeV hadron collider with an uncertainty on the background determination $\epsilon_{syst}=20\%$. Right panel: Projected 5$\sigma$ discovery (blue) and $2\sigma$ discrimination (gray) reach, using the $p_T$ distributionof the sub-leading lepton with a binning of 300 GeV.}
\end{figure}
Furthermore we assume that the number of signal events necessary for a 2$\sigma$ and 5$\sigma$ significance are equal with respect to the ones obtained for the 33 TeV case for the same value of integrated luminosity. While these could be considered strong assumptions they avoid the need of simulating background events, and we believe them to be reasonable due to the very preliminary nature of such projections. Note however that the assumption of keeping a constant selection efficiency for the signal does not affect in principle the shape of any distribution but only their relative rescalings between the different hypotheses.
It therefore allows us to evaluate the pure effect of shape differences between left-handed and right-handed chirality assumptions; as we have observed a mild dependence on the differences between selection efficiencies at 13 TeV and 33 TeV, we implicitly assume that a similar effect will take place also at 100 TeV.
In the left panel of Fig.~\ref{fig:100Lumi} we show the exclusion and discovery reaches on the $X$ mass, while in the right panel we compare the discovery reach with the 2$\sigma$ discrimination reach, considering again the distribution of the transverse momentum of the second leading lepton with a binning of 300 GeV. Also in this case our estimate shows that the discrimination among the two coupling hypotheses its possible in all the discovery range accessible at a 100 TeV collider and for a VLQ with mass greater than $\sim 4300\;$GeV, which is discoverable with $\sim 800\;$fb$^{-1}$ of integrated luminosity, the discrimination will be possible with the already collected data set.

\section{Conclusions}
\label{sec:conc}
In this paper we have discussed how to disentangle the dominant coupling chirality of VLQs which interact with the SM top quark if they are observed at the LHC or future higher-energy colliders. The coupling chirality of a VLQ is related to its quantum number under the $SU(2)_L$ SM gauge group. Representations with odd weak isospin have dominantly left-handed couplings, while those with even weak isospin are dominantly right-handed. Therefore the determination of the coupling structure would provide an essential piece of information for embedding VLQs into theoretically motivated scenarios. 

We have firstly evaluated the degree of polarisation of the gauge bosons emerging from a VLQ with charge 2/3 decaying to $Wb$ or $Zt$. For VLQ masses still allowed by current bounds, around 1 TeV, the degree of transverse polarisation is always small, of the order of 1\%, thus making its observation rather challenging.
However in the case that VLQs decay to top quarks, they will induce a preferred polarisation for the top which depend on the dominant chirality of the coupling of the VLQ. This in turn affects the kinematic properties of the top quark decay products. In particular we have exploited the differences of the leptons arising from a leptonic top quark decay to perform a shape analysis and compare the 2$\sigma$ discrimination reach of the LHC between left- and right-handed coupling hypotheses  obtained through a $\chi^2$ analysis of the lepton $p_T$ distributions  with the discovery reach of an experimental search for $T$ VLQ decaying into $Zt$ for various values of integrated luminosity.
Our results show that if a $T$ with mass $\gtrsim$ 1200 GeV and decaying to $Zt$ is discovered at the LHC in the single lepton channel, it will be possible to identify its dominant coupling chirality at 95\% CL with luminosities of at least 300 fb$^{-1}$.

Due to the mild increase in the mass discovery reach which is expected with higher LHC integrated luminosities it is crucial to determine the potentiality of future proposed collider in performing a similar analysis. We have considered two possible future high-energy proton-proton colliders with a centre of mass energy of 33 TeV and 100 TeV and considered a process of pair production of a VLQ with charge 5/3 decaying entirely to $Wt$. In this case, we have verified that the kinematics of the process allows us to identify with very good accuracy the second leading lepton with the lepton coming from the decay of the SM top quark. We have therefore considered a signal region in the same-sign dilepton channel and performed again $\chi^2$ analysis for both centre of mass hypotheses. We found that at 33 TeV it will be possible to discriminate the coupling chirality of $X$ VLQs with mass above $\sim$2300 GeV with luminosities larger than $\sim$1 ab$^{-1}$, while at 100 TeV the discrimination can be obtained for masses above $\sim$4.3 TeV and luminosities above
$\sim$800 fb$^{-1}$. Interestingly, the results for $X$ can be reinterpreted in terms of a charge 2/3 $B$ VLQ decaying entirely into $Wt$, as the kinematics of the final state still implies that the second leading lepton arises from the decay of a top quark.\\

To summarise, the characterisation of the coupling properties of VLQs would be extremely useful from a theoretical point of view in the case of a future discovery, and with this exploratory analysis we have shown that such characterisation is experimentally possible, at least for scenarios where the VLQs decay into SM top quarks. The characterisation can be obtained through a simple shape analysis of the leptons arising from the top quark decay and we have proven that positive results can be obtained already with selection and kinematic cuts used in current experimental searches, therefore making the potentialities for simultaneous discovery and characterisation at the LHC and future colliders very promising.

\section*{Acknowledgements}
The authors wish to thank S. Tosi for useful discussions, and acknowledge the use of the IRIDIS HPC Facility at the University of Southampton for the completion of this study.

\section*{Note Added}
Further results with projections for the 27 TeV LHC and a more detailed study of the background separation can be found in the Report of the Physics of the HL-LHC and Perspectives at the HE-LHC, Working Group 3, Beyond the Standard Model Physics~\cite{CidVidal:2018eel}.

\bibliography{refs.bib}
\bibliographystyle{JHEP}

\end{document}